\documentclass[12pt]{article}

\usepackage{amsmath}
\usepackage{newtxtext,newtxmath}

\usepackage{graphicx}

\usepackage{array}

\usepackage[letterpaper,left=1in,right=1in,top=0.85in,bottom=0.85in]{geometry}

\makeatletter
\renewcommand{\fps@figure}{htbp}
\renewcommand{\fps@table}{htbp}
\makeatother

\renewenvironment{abstract}
	{\quotation\vspace{-1em}}
	{\endquotation}

\date{}

\usepackage{titling}
\makeatletter
\renewcommand{\fnum@figure}{\textbf{Figure \thefigure}}
\renewcommand{\fnum@table}{\textbf{Table \thetable}}
\makeatother

\usepackage{url}

\usepackage{bm}          
\usepackage{booktabs}    
\usepackage{float}       
\usepackage{enumitem}    
\usepackage{subcaption}  
\usepackage{tikz}        
\usetikzlibrary{positioning, arrows.meta, calc, fit, backgrounds, decorations.pathreplacing}
\usepackage[round,authoryear]{natbib} 

\newcommand{\FIGURE}[3]{%
  \centering
  #1\par\smallskip
  \caption{\textbf{#2} #3}%
}
\newcommand{\TABLE}[3]{%
  \centering
  \caption{\textbf{#1}}%
  \par\smallskip
  #2\par
  \ifx\relax#3\relax\else\smallskip{\small\leftskip=0pt\rightskip=0pt\parfillskip=0pt plus 1fil #3\par}\fi
}
\providecommand{\up}{\rule{0pt}{2.6ex}}
\providecommand{\down}{\rule[-1.2ex]{0pt}{0pt}}

\providecommand{\NOTE}[1]{\textit{#1}\ \ignorespaces}

\def\scititle{
	When and How Should a Power Trader Engage in Arbitrage? Predict, then Contextually Optimize
}
\title{\bfseries \boldmath \scititle}

\author{
    Yannick Heiser$^{1}$, Jalal Kazempour$^{1}$, Farzaneh Pourahmadi$^{1}$\and
	\small$^{1}$Department of Wind and Energy Systems,
Technical University of Denmark.\and\
	\small$^{1}$Email: \{yahei, jalal, farpour\}@dtu.dk
}

\begin{document} 

\maketitle

\begin{abstract} \bfseries \boldmath
Electricity markets increasingly expose stochastic energy generators to arbitrage opportunities between the day-ahead and balancing markets, driven by widening price spreads. However, opportunistic bidding, intentionally deviating from the production forecast to exploit anticipated price spreads, carries significant risk, and existing frameworks rarely offer explainable, risk-aware decision support. We propose a \textit{predict-then-contextual-optimize} framework that decomposes the day-ahead bidding decision into three explicit stages to decide when to engage in arbitrage, in what direction, and to what extent. A probabilistic binary classifier with confidence thresholds determines whether the prediction of the price spread is sufficiently confident to justify an opportunistic bid. Otherwise, the trader defaults to an arbitrage-free bid equal to the power forecast. A linear decision policy learned for each class via contextual optimization determines the magnitude of the bid deviation from the power forecast. The framework accommodates both standalone renewable generation and hybrid power plants combining renewable generation with other assets, such as an electrolyzer. We evaluate the framework on a real wind farm in the European bidding zones DK1 and DE/LU using a rolling-window procedure and compare it against several benchmark bidding strategies. The results show that the proposed framework increases mean profit relative to an arbitrage-free benchmark, reaching an improvement of about 7\% for the hybrid power plant in DK1. The largest gains occur when distributional drift between training and testing windows is low, while the co-located electrolyzer further increases arbitrage value by providing additional operational flexibility.
\end{abstract}

\section{Introduction}\label{sec:Intro}
A stochastic energy generator, e.g., a wind farm, typically sells a majority of its energy for a delivery period $t$ in the day-ahead electricity market at price $\lambda_t^{\rm DA}$ (€/MWh). One day prior to physical delivery $t$, it needs to decide on the amount of energy $p_t^{\rm DA}$ (MWh) to bid into the day-ahead market, given a production forecast $\hat{P}_t^{\rm W}$ (MWh) and assuming a marginal production cost of €0/MWh. Given the realized production $P_t^{\rm W}$ after the delivery period $t$, any physical deviation from the contracted energy $p_t^{\rm B} = P_t^{\rm W} - p_t^{\rm DA}$ is settled post-delivery in the balancing market at price $\lambda_t^{\rm B}$ (€/MWh), which may be lower than, equal to, or higher than $\lambda_t^{\rm DA}$, depending on the system need, i.e., whether there is a power deficit or surplus in the system. If the trader bids its production forecast $\hat{P}_t^{\rm W}$, we refer to this as an \textit{arbitrage-free bid}, since the trader does not intentionally exploit a forecasted price difference between the two markets. By contrast, if the trader intentionally bids above the forecast to benefit from a positive price spread $\Delta\lambda_t = \lambda_t^{\rm DA} - \lambda_t^{\rm B}$, we call this an \textit{opportunistic long arbitrage bid}. If the trader bids below the forecast to benefit from a negative price spread, we call this an \textit{opportunistic short arbitrage bid}.

However, both market prices $(\lambda_t^{\rm DA}, \lambda_t^{\rm B})$ are uncertain at the time of decision making in the day-ahead stage, introducing substantial risk to opportunistic bidding strategies. Recent market developments further motivate this problem, as increasing renewable penetration, higher balancing-price volatility, and ongoing changes in balancing market design make arbitrage opportunities between the day-ahead and balancing markets more relevant. At the same time, these trends also make opportunistic arbitrage decisions riskier and more difficult. Therefore, an effective trading strategy needs to account for the uncertainty of $\Delta\lambda_t$ and explicitly consider the risk of each decision. In addition, the trading framework should be pragmatic for power traders, i.e., computationally efficient and explainable. The problem becomes even more challenging for co-located assets with temporal constraints, such as a hybrid power plant (HPP) combining renewable generation and electrolyzer operation. In this work, we focus solely on the perspective of a price-taking power trader and leave the system-level implications of opportunistic arbitrage bidding, such as effects on market social welfare, for future work.

\subsection{Literature Review}

Literature on bidding stochastic renewable energy into the day-ahead market is mostly developed under \textit{dual-price} balancing schemes, where separate imbalance prices apply depending on whether the power trader's imbalance supports or aggravates the system need. In this setting, the trader faces different prices for upward and downward deviations, and any imbalance is typically penalized relative to the day-ahead position. The resulting bidding problem is therefore commonly formulated as a classical newsvendor-type problem  \citep{Pinson2023DistributionallyProducers,Aolaritei2025HedgingMarkets}. Under such a dual-price scheme, the trader is primarily exposed to the risk of imbalance costs, while intentional arbitrage between the day-ahead and balancing markets is structurally limited. By contrast, several European markets have recently moved toward harmonized imbalance settlement rules, including \textit{single-price} balancing, where a balance responsible party is settled at one balancing price irrespective of the direction of its imbalance \citep{ACER2024ImbalanceSettlementHarmonisation,ACER2020ImbalanceSettlementMethodology}. While this imbalance settlement mechanism exposes stochastic producers to price risk, it can also create opportunities for arbitrage between the day-ahead and balancing markets. We describe this market design in more detail in Section~2. Despite this practical relevance, there is still limited literature on opportunistic arbitrage bidding for stochastic producers under single-price balancing.

A structurally related problem has been studied more extensively in U.S. two-settlement electricity markets under the terms \textit{virtual bidding} or \textit{convergence bidding}. Virtual bidding allows purely financial participants to take positions in the day-ahead market and close them in the real-time market without holding physical generation or consumption assets. In this setting, a trader can submit an increment or decrement bid in the day-ahead market and offset this position in real time, thereby arbitraging the price spread between the two settlements. Several studies analyze virtual bidding from a market design perspective, showing that it can improve liquidity, price convergence, and market efficiency under suitable conditions \citep{Isemonger2006BenefitsRisksVirtualBidding,Hogan2016VirtualBiddingElectricityMarketDesign,JhaWolak2023ForwardCommodityMarkets}. However, other works show that these benefits may not materialize when virtual bids interact with non-convex market-clearing procedures, transmission constraints, or financial transmission rights, and that virtual bidding may also create incentives for market manipulation \citep{Parsons2015FinancialArbitrage,Ledgerwood2013UsingVirtualBids,Hopkins2020ConvergenceBids}. These studies are important for understanding the market-level implications of arbitrage, but they differ from our focus. In this paper, we take the perspective of a price-taking power trader and study how such a trader should make risk-aware arbitrage decisions. We do not analyze the welfare, liquidity, or price convergence impacts of opportunistic arbitrage bidding, which we leave for future work.

Within the literature taking a power trader's perspective, several papers propose optimization or learning methods for virtual trading.~\citet{Xiao2018} develop stochastic optimization models for increment and decrement bidding curves using generated price scenarios and conditional value-at-risk (CVaR) constraints for risk management.~\citet{Baltaoglu2019Algorithmic} study algorithmic virtual bidding from an online learning perspective and propose a dynamic programming method that sequentially allocates a fixed budget across multiple location-hour opportunities without assuming knowledge of the underlying price distribution. More recent learning-based approaches also use historical data to predict or exploit day-ahead and real-time price spreads \citep{Li2022MachineLearningVirtualBidding}. These papers demonstrate the value of data-driven and risk-aware methods for arbitrage bidding. However, they are primarily designed for financial traders in U.S. markets and do not consider the physical constraints of renewable generators or co-located assets. Moreover, the link between observable market conditions, price spread uncertainty, and the final arbitrage decision is often not explainable. Although the models may use historical spread data, they do not explicitly explain when arbitrage should be avoided, what drives the predicted arbitrage direction, or how a trader can tune its risk exposure in a practically explainable way.

The literature on European single-price balancing markets for stochastic assets is more limited.~\citet{Browell2018Risk} studies risk-constrained trading strategies for stochastic generation under a single-price balancing market by adapting the \textit{all-or-nothing} strategy, which will be discussed in Section~\ref{sec:overview}, using system imbalance state predictions. However, predicting the system imbalance state may be insufficient when the relevant economic quantity is the price spread between the day-ahead and balancing markets. \citet{Bruninx2025} introduce a risk certificate that bounds the balancing market position and thereby restricts the all-or-nothing bid. This provides a useful way to limit exposure, but the balancing price, which is the main source of uncertainty in the arbitrage decision, is represented by a simplified rolling average. \citet{Heiser2025} learn linear decision policies that directly map contextual features to decisions for wind and hydrogen trading under risk constraints such as CVaR. This is a form of decision rule optimization in the sense of Ban and Rudin (2019) extended to the risk-constrained setting. Although computationally efficient, the prediction of the price spread is embedded implicitly in the policy learning step rather than being made explicit to the trader. 

Our work builds upon the broader literature on data-driven decision-making, particularly contextual optimization. As reviewed in~\citet{Sadana2024Survey}, contextual optimization methods can be broadly categorized into three approaches. In predict-then-optimize, a predictive model is trained independently of the downstream decision problem, and its forecasts are subsequently used in optimization~\citep{Bertsimas2020PredictivePrescriptive}. In decision-focused learning, the predictive model is trained by explicitly accounting for the downstream decision loss~\citep{Donti2017,Elmachtoub2020SmartOptimize,Mandi2024}.
Within energy applications, decision-focused approaches have also been developed under the term value-oriented forecasting, particularly for renewable energy trading~\citep{Carriere2019,Stratigakos2022} and energy storage arbitrage~\citep{Sang2022}.
In decision rule optimization, also referred to as direct policy learning, a parameterized policy maps contextual features directly to decisions and is learned by optimizing the empirical decision objective~\citep{Ban2019}. Our proposed framework combines elements of predict-then-optimize and decision rule optimization. In the first step, a classifier predicts the direction of the electricity price spread, while the decision to engage in arbitrage is determined through thresholds tuned according to downstream economic performance. In the second step, class-specific linear decision policies are learned through contextual optimization to determine the magnitude of the arbitrage position while accounting for the trader's risk preferences. The novelty lies in how the reject option is embedded in the decision pipeline: its thresholds are tuned on the downstream profit, and the samples it selects define the training data of the class-specific policies in the second step.

The confidence-threshold mechanism in the first step is closely related to selective classification, also known as classification with a reject option, in which a classifier abstains from making predictions when its confidence is insufficient~\citep{Hendrickx2024}. This concept was initially formalized by~\citet{Chow1970} through a trade-off between misclassification and rejection costs. In their cost-based formulation, each rejection incurs a fixed cost and the optimal rule rejects a sample whenever its highest posterior class probability falls below a threshold determined by this cost. Subsequent studies have investigated alternative formulations that maximize prediction coverage subject to a bound on selective risk~\citep{ElYaniv2010, Geifman2017}, or minimize selective risk subject to a target coverage~\citep{Franc2023}. Rejection mechanisms can be constructed by applying thresholds to a pretrained classifier, as in our framework, or learned jointly with the classifier~\citep{Cortes2016}. Moreover, asymmetric rejection thresholds have been studied when misclassification and rejection costs differ between classes as in our case~\citep{Pietraszek2005}. Building on this literature, our framework incorporates a rejection mechanism into the arbitrage bidding decision. Unlike conventional selective classification approaches that typically focus on classification errors and coverage, our rejection thresholds are tuned based on downstream economic performance, including expected profit or CVaR.

Across these streams, two research gaps remain. First, existing arbitrage-bidding models \citep{Xiao2018,Baltaoglu2019Algorithmic,Li2022MachineLearningVirtualBidding,Browell2018Risk,Bruninx2025,Heiser2025} rarely provide an explainable decomposition of the decision into \textit{when} to engage in arbitrage, \textit{which direction} to take, and \textit{how large} the position should be. In principle, the models in~\citet{Xiao2018,Parsons2015FinancialArbitrage} and~\citet{Avila2025Multistage} could merge these layers and optimize a single arbitrage quantity directly, where zero means no arbitrage, a positive value means a long position, and a negative value means a short position. However, such an integrated formulation obscures the economic logic of the decision and may require solving an optimization problem even in situations where the price spread signal is too uncertain and arbitrage should be avoided. 
Second, existing approaches offer limited practical tools for managing the trader's risk preference. Some models are risk-neutral \citep{Bathurst2002, Pinson2007}, while others impose risk aversion through CVaR optimization \citep{Rockafellar2000} or scenario-based formulations \citep{Morales2010,Baringo2016}. In many cases, however, it is unclear how a trader can tune the degree of opportunism in deployment and understand the resulting profit--risk trade-off. 

\subsection{Contributions}
As mentioned above, the existing bidding strategies  mainly focus on how to trade in the market, often producing arbitrage bids implicitly as the output of an optimization or learning model. However, this makes it difficult for the trader to understand when arbitrage is justified and when it should be avoided. 
To overcome this issue, our framework combines the logic of selective classification with predict-then-optimize and contextual optimization. In the first step, we predict the sign of the price spread $\bm{\Delta\lambda}$ from contextual features. 
The output is turned into decisions by a reject option in the sense of selective classification, whose thresholds are tuned on the mean or the CVaR of the downstream profit.
In the second step, we use contextual optimization to learn bidding policies that map the same contextual features directly to day-ahead bid quantities. 
We therefore refer to the proposed methodology as a \textit{predict-then-contextual-optimize} framework. This positioning connects our approach to the selective prediction literature \citep{Hendrickx2024}, the predict-then-optimize literature \citep{Elmachtoub2020SmartOptimize} and the contextual optimization literature \citep{Bertsimas2020PredictivePrescriptive,Sadana2024Survey}, while preserving the explainability needed for practical bidding decisions. The proposed framework explicitly decomposes the trading decision into three stages, as illustrated in Figure~\ref{fig:decision framework}:
\begin{figure}
     \FIGURE
{
\begin{tikzpicture}[
    x=0.813cm, y=1cm,
    >=stealth,
    every node/.style={outer sep=0pt},
    decbox/.style={
        rectangle, draw=blue!60!black, fill=blue!8,
        rounded corners=3pt,
        minimum width=2.4cm, minimum height=1.0cm,
        align=center, font=\fontsize{10}{12}\selectfont
    },
    mbox/.style={
        rectangle, draw=green!60!black, fill=green!8,
        rounded corners=3pt,
        minimum width=2.4cm, minimum height=1cm,
        align=center, font=\fontsize{10}{12}\selectfont
    },
    ebox/.style={
        rectangle, draw=orange!70!black, fill=orange!8,
        rounded corners=3pt,
        minimum width=2.4cm, minimum height=1.0cm,
        align=center, font=\fontsize{10}{12}\selectfont
    },
    slbl/.style={font=\fontsize{10}{12}\selectfont\bfseries, align=center},
    plbl/.style={align=center, font=\fontsize{10}{12}\selectfont},
    bglayer/.style={draw=gray!45, fill=gray!10, rounded corners=6pt},
    llbl/.style={align=center, inner sep=4pt, text=black!75, font=\fontsize{9}{11}\selectfont\itshape}
]
\node[slbl, text=black]  at (0, 2.85) {\textit{When?}};
\node[slbl, text=black]  at (4.5, 2.85) {\textit{What direction?}};
\node[slbl, text=black]  at (9, 2.85) {\textit{What extent?}};
\draw[decorate, decoration={brace, amplitude=6pt}, thick, black]
    (4.5, 3.2) -- (9, 3.2)
    node[midway, above=6pt, slbl, text=black] {\textit{How?}};
\node[plbl] (CTX) at (-3.6, 0) {Context};
\node[mbox] (W) at (0, 0) {Engage in\\[-2pt]arbitrage?};
\draw[->, semithick] (CTX.east) -- (W.west);
\node[plbl] (NO) at (0, -2.7) {Arbitrage-free bid};
\draw[->, semithick, black] (W.south) --
    node[right, font=\fontsize{10}{12}\selectfont] {No} (NO.north);
\node[mbox] (M) at (4.5, 0) {Direction of\\[-2pt]arbitrage};
\draw[->, semithick, green!50!black] (W.east) --
    node[above, font=\fontsize{10}{12}\selectfont] {Yes} (M.west);
\node[decbox] (EXTL) at (9,  1.4) {Choose risk\\[-2pt]level};
\node[decbox] (EXTS) at (9, -1.4) {Choose risk\\[-2pt]level};
\draw[->, semithick, green!50!black] (M.east) -- (EXTL.west)
    node[pos=0.5, xshift=-14pt, above, font=\fontsize{10}{12}\selectfont, text=green!50!black] (LONG) {Long};
\draw[->, semithick, green!50!black] (M.east) -- (EXTS.west)
    node[pos=0.5, xshift=-14pt, below, font=\fontsize{10}{12}\selectfont, text=green!50!black] (SHORT) {Short};
\node[plbl] (OUTL) at (13.5,  1.4) {Opportunistic long bid};
\node[plbl] (OUTS) at (13.5, -1.4) {Opportunistic short bid};
\draw[->, semithick] (EXTL.east) -- (OUTL.west);
\draw[->, semithick] (EXTS.east) -- (OUTS.west);
\node[inner sep=6pt, fit=(W)(M)(LONG)(SHORT)(EXTL.north -| W)(EXTS.south -| W)] (L1) {};
\node[inner sep=6pt, fit=(EXTL)(EXTS)] (L2) {};
\node[llbl, anchor=north east] (T1) at ([xshift=-6pt]L1.south east) {Confidence analysis\\[-2pt]on system status};
\node[llbl, anchor=north] (T2) at (L2.south) {Individual\\[-2pt]decision making};
\begin{scope}[on background layer]
\foreach \t in {T1, T2}{%
    \draw[bglayer] ([yshift=6pt]\t.north west) |- (\t.south east) -- ([yshift=6pt]\t.north east);
}
\draw[bglayer] (L1.south west) rectangle (L1.north east);
\draw[bglayer] (L2.south west) rectangle (L2.north east);
\foreach \x in {0, 4.5, 9}{%
    \draw[dashed, gray!55, thin] (\x,0 |- L1.south) -- (\x, 2.7);
}
\end{scope}
\node[draw=black!40, fill=white, rounded corners=3pt,
      inner sep=4pt, anchor=south west]
    (legend) at (10.5, 2.6) {%
    \begin{tabular}{@{}cl@{}}
        \raisebox{-0.05cm}{\tikz\draw[mbox, minimum width=0.35cm, minimum height=0.35cm] (0,0) rectangle (0.35,0.35);} &
        \fontsize{10}{12}\selectfont Prediction \\[-4pt]
        \raisebox{-0.05cm}{\tikz\draw[decbox, minimum width=0.35cm, minimum height=0.35cm] (0,0) rectangle (0.35,0.35);} &
        \fontsize{10}{12}\selectfont Contextual Optimization
    \end{tabular}%
};
\end{tikzpicture}
}
{Arbitrage bidding framework.\label{fig:decision framework}}
{Given context as input, the proposed framework determines when to engage in arbitrage, in what direction, and to what extent. The colors refer to the different model parts that will be introduced in Section~\ref{sec:train_valid}: a probabilistic classification model for prediction (green) and two linear policies learned using contextual optimization (blue). For a given sample, the framework can have either an arbitrage-free bid or an opportunistic long or short bid as output. \color{black} The gray background layers group the stages into a confidence analysis on the system status, which is common to all market participants, and the individual decision making of the trader.\color{black}}
\end{figure}
\begin{enumerate}[leftmargin=*, widest=3.]
\item \textbf{When to bid opportunistically.} We introduce a probabilistic binary classifier coupled with direction-specific \textit{confidence thresholds} on the predicted sign of the price spread $\Delta\lambda_t$ to identify when opportunistic bidding is justified. When the prediction is too uncertain, the framework defaults to placing an arbitrage-free bid, i.e., day-ahead bid equal to the power forecast. We refer to this as a confidence analysis on the system status, which is identical for all market participants, as it is based on the predicted sign of the price spread.

\item \textbf{Direction of the opportunistic bid.} Conditional on engaging in arbitrage, the framework determines whether the trader should take a long or short position relative to the production forecast. A predicted positive price spread $\Delta\lambda_t$ motivates bidding above the forecast (opportunistic long bid), while a predicted negative price spread motivates bidding below the forecast (opportunistic short bid).

\item \textbf{Extent of the position.} The magnitude of the long or short position is determined based on a linear decision policy per class that is trained by maximizing a weighted combination of expected profit and CVaR of the profit via contextual optimization. This provides a second layer of risk management in addition to the confidence thresholds. We refer to this stage as individual decision making, as it is based on the trader's own risk preference and can be different for each market participant.

\end{enumerate}

This risk-aware structure of three stages preserves practicality, enables fast decision-making at testing time, and allows operational constraints, such as those of a co-located electrolyzer in an HPP, to be incorporated directly into policy training without structural modification of the framework.
In addition, we present a realistic  case study demonstrating the framework on a real wind farm in the DK1 and DE/LU bidding zones, comparing models of varying complexity and quantifying the profit--risk trade-off introduced by the confidence thresholds.
 
The remainder of the paper is structured as follows: Section~\ref{sec:preliminaries_overview} provides market preliminaries and a detailed problem overview.  Section~\ref{sec:train_valid} details the training phase, covering both the classification (Section~\ref{sec:classif}) and the contextual optimization problem (Section~\ref{sec:optimization}). Section~\ref{sec:testing} presents the full testing phase, describing how a trained model processes contextual features to determine optimal day-ahead bids and compute realized profits, whereas Section~\ref{sec:Results} provides numerical results. Finally, Section~\ref{sec:Conclusion} concludes the paper.

\section{Preliminaries and Overview}\label{sec:preliminaries_overview}
This section establishes the foundations for the bidding framework developed in the paper. Section~\ref{sec:preliminaries} introduces the market structure and motivates the problem by characterizing the statistical properties of the day-ahead ($\bm{\lambda}^{\rm DA}$) and balancing prices ($\bm{\lambda}^{\rm B}$) over the previous years.  Building on this, Section~\ref{sec:overview} formalizes the problem setting, presents the overview of the three-stage decision framework, and describes the rolling window procedure used throughout the paper to evaluate the out-of-sample performance.
\subsection{Market-Related Preliminaries} \label{sec:preliminaries}
As introduced in Section~\ref{sec:Intro}, we focus on a two-stage bidding problem. Specifically, the first trading floor is the day-ahead market, in which bids are placed on the day before, $d-1$, for delivery periods on day $d$. Hence, all bids of delivery periods of day $d$ need to be submitted before gate closure on $d-1$. In Europe, the day-ahead market has gate closure at noon on $d-1$ and the delivery periods have recently transitioned from a 60-minute market time unit (MTU) to a 15-minute MTU, quadrupling the number of products to 96 for delivery day $d$. Therefore, the market clearing yields day-ahead prices $\bm{\lambda}^{\rm DA}_d \in \mathbb{R}^{96}$ for each delivery day $d$. The second trading floor is the balancing market, in which any imbalance from previous floors is automatically settled ex-post after realization. In Europe, the balancing market has also recently moved from a 60-minute to a 15-minute MTU, yielding balancing prices $\bm{\lambda}^{\rm B}_d \in \mathbb{R}^{96}$ for day $d$. As the MTU is reduced, the power system can switch more frequently between surplus and deficit conditions within the delivery day. These more frequent changes in the system imbalance condition can increase balancing price volatility and enlarge the price spread between the day-ahead and balancing markets. Consequently, arbitrage becomes more relevant, as traders have more opportunities to benefit from deviations between the two market prices.
\begin{figure}
    \FIGURE
    {%
        \begin{minipage}[t]{\textwidth}
            \centering
            \input{prices_DK1.pgf}
        \end{minipage}
    }%
    {Day-ahead and balancing prices and their spreads for DK1.\label{fig:prices_DK1}}
    {(a) Time series of the day-ahead prices (red) and balancing prices (blue) for the Danish bidding zone DK1. The vertical dashed line marks the start of the new balancing reserve activation method in March 2025, after which the balancing prices exhibit a clear shift in their dynamics. The period after this shift is the focus area of our numerical case study. (b) Distribution of the price spread $\bm{\Delta\lambda}$ over the focus period, shown on a logarithmic scale. The vertical lines denote the mean (dotted) and median (dashed) values of the distribution.}
\end{figure}

In Figure~\ref{fig:prices_DK1}a, the day-ahead prices (red) and balancing prices (blue) are shown for the Danish bidding zone DK1 from January 2023 to February 2026. The balancing prices exhibit a clear shift in dynamics, caused by the new activation method for tertiary balancing reserves \citep{NBM2026}. We focus on the period after this shift, as it is characterized by much higher price spreads, and therefore, greater arbitrage opportunities. The distribution of the price spreads $\bm{\Delta\lambda}$ over the focus period is shown in Figure~\ref{fig:prices_DK1}b on a logarithmic scale. While most samples have a positive spread (median price spread = €16.7/MWh), the negative spread has a much heavier tail pushing the distribution's mean ($-$€3.3/MWh) well below its median. This tail is mainly driven by the high balancing price peaks that are also visible in Figure~\ref{fig:prices_DK1}a.

\subsection{Problem Overview}\label{sec:overview}
We consider a price-taking power trader, either a standalone wind farm (\textit{wind-only)} or an HPP consisting of a co-located wind farm and electrolyzer that participates in two sequential electricity markets, the day-ahead market and the balancing market. Recall that the power trader submits the day-ahead bid $p_t^{\rm DA}$ before the wind power production $P_t^{\rm W}$, day-ahead and balancing prices, $\lambda_t^{\rm DA}$ and $\lambda_t^{\rm B}$, are realized. 
Under the current market regulation, the optimal bidding strategy in hindsight, i.e., with perfect information on prices and production, follows a binary \textit{all-or-nothing} rule. To see this, consider the profit-maximization problem for a wind farm over a single delivery period $t$:
\begin{subequations}\label{eq:windfarm}
\begin{align}
    \underset{p_t^{\rm DA},p_t^{\rm B}}{\max} \quad & \lambda^{\rm DA}_t p_t^{\rm DA} + \lambda^{\rm B}_t p_t^{\rm B} \label{eq:windfarm_obj} \\
    \text{s.t.} \quad & f(p_t^{\rm DA}, p_t^{\rm B}) \leq 0,\label{eq:windfarm_constraint}
\end{align}
\end{subequations}
where $f(p_t^{\rm DA}, p_t^{\rm B}) \leq 0$ represents the power balance constraint 
$P^{\rm W}_t = p_t^{\rm DA} + p_t^{\rm B}$ and the day-ahead quantity bid bounds 
$0 \leq p_t^{\rm DA} \leq \overline{P}^{\rm W}$. Substituting the power balance into the objective function and defining $\Delta\lambda_t=\lambda^{\rm DA}_t-\lambda^{\rm B}_t$ yields:
\begin{equation*}
    \lambda^{\rm DA}_t p_t^{\rm DA} + \lambda^{\rm B}_t \bigl(P^{\rm W}_t - p_t^{\rm DA}\bigr) = \Delta\lambda_t p_t^{\rm DA} + \lambda^{\rm B}_t P^{\rm W}_t.
\end{equation*}

Since $\lambda^{\rm B}_t P^{\rm W}_t$ is constant in $p_t^{\rm DA}$, the problem reduces to maximizing $\Delta\lambda_t p_t^{\rm DA}$ over $[0, \overline{P}^{\rm W}]$. As the objective is linear, the optimum is always attained at a boundary:
\begin{itemize}
    \item If $\Delta\lambda_t > 0$: $p_t^{\rm DA\,*} = \overline{P}^{\rm W}$, i.e., bid the full wind capacity in the day-ahead market.
    \item If $\Delta\lambda_t < 0$: $p_t^{\rm DA\,*} = 0$, i.e., withhold all energy from the day-ahead market.
\end{itemize}

Hence, with perfect information the optimal strategy depends solely on $\text{sign}(\Delta\lambda_t)$. Note that problem~\eqref{eq:windfarm} is not solved or used as a training objective in our framework. It only identifies $\text{sign}(\Delta\lambda_t)$ as the natural prediction target. Under uncertainty, predicting this sign is no longer sufficient, since the trader must also decide how far to deviate from the arbitrage-free bid given the risk that the prediction is wrong.

This result extends to the HPP case of a co-located wind farm 
and electrolyzer. Consider the analogous profit-maximization problem:
\begin{subequations}\label{eq:hpp}
\begin{align}
    \underset{p_t^{\rm DA},p_t^{\rm B},h_t}{\max} \quad & \lambda^{\rm DA}_t p_t^{\rm DA} + \lambda^{\rm B}_t p_t^{\rm B}  + \lambda^{\rm H}h_t\label{eq:hpp_obj} \\
    \text{s.t.} \quad & g(p_t^{\rm DA}, p_t^{\rm B}, h_t) \leq 0,\label{eq:hpp_constraint}
\end{align}
\end{subequations}

\noindent where $g(p_t^{\rm DA}, p_t^{\rm B}, h_t) \leq 0$ represents the power balance 
and operational constraints of the HPP, with $h_t$ and $\lambda^{\rm H}$ denoting the production and price of hydrogen, respectively. 
\citet{Heiser2025} show that the binary all-or-nothing bidding rule in hindsight holds numerically for the HPP setting as well. 

As established above, the sign of the price spread 
$\Delta\lambda_t$ is the determinant of the optimal bidding decision, yet it is inherently difficult to forecast due to the volatile balancing prices. This forecasting uncertainty introduces substantial decision risk, motivating the need for a bidding framework that can exploit arbitrage opportunities while considering the price uncertainty. Hence, a central challenge is to learn when and how to deviate from an arbitrage-free bid in a risk-aware and data-driven manner.

Our predict-then-contextual-optimize framework addresses this challenge through two coupled modeling steps. Step 1 is a probabilistic binary classifier that maps the contextual features $\mathbf{x}_t$ available at the day-ahead bidding time stage to a score in $[0,1]$, which shows how strongly the features indicate a positive ($\Delta\lambda_t > 0$) or negative price spread ($\Delta\lambda_t < 0$). Two tunable confidence thresholds convert the output into one of the three arbitrage trading decisions, i.e., they yield either an opportunistic long or short or arbitrage-free bid. Step 2 is a contextual optimization model that learns two linear decision policies, one for going long and one for going short, by directly mapping features to bid quantities. The two steps answer different questions and are trained with different objectives. Step~1 is a confidence analysis on the system status, i.e., whether the day-ahead price is higher or lower than the balancing price, a question that is the same for every market participant. Step~2 is the individual decision of the trader, which depends on its individual risk preference. We assign the decision of when and in which direction to engage in arbitrage to Step~1 and only the extent to Step~2, as we aim to engage in arbitrage only if we can make a sufficiently confident prediction about the system status. The relation between the two steps is discussed in Section~\ref{sec:train_valid}, and the profit calculation for the testing phase (Step 3) is described in Section~\ref{sec:testing}.

\paragraph{Notation.}
Throughout the paper, we use $t \in \mathcal{T}^{\rm train}$ to index time periods in the training set, $\kappa \in \mathcal{T}^{\rm val}$ to index time periods in the validation set, and $\tau \in \mathcal{T}^{\rm test}$ to index time periods in the testing set. Realized quantities are written without decoration (e.g.,\ $\lambda_t^{\rm B}$ and $P_t^{\rm W}$), forecasts are denoted by a hat (e.g.,\ $\hat{P}_t^{\rm W}$), and learned or tuned parameters are denoted by a tilde when evaluated in testing (e.g.,\ $\tilde{\bm{\theta}}$ and $\tilde{\overline{\alpha}}$).

\paragraph{Rolling Windows.}
The full framework is trained and evaluated via a rolling window procedure, illustrated in Figure~\ref{fig:rolling_windows}. The full data is split into different windows \{$w1$, $w2$, ...\}, which typically span over several months. Each window $w$ consists of three non-overlapping, chronologically ordered sets as follows:
\definecolor{traincolor}{RGB}{59, 130, 246}
\definecolor{validcolor}{RGB}{245, 158, 11}
\definecolor{testcolor}{RGB}{16, 185, 129}

\begin{figure}
     \FIGURE
{
\begin{tikzpicture}[
    lbl/.style={font=\fontsize{10}{14}\selectfont\sffamily, text=white},
    rowlbl/.style={font=\fontsize{10}{14}\selectfont\sffamily, anchor=east},
    axislbl/.style={font=\fontsize{10}{14}\selectfont\sffamily, text=gray!70},
    every node/.style={outer sep=0},
]
\pgfmathsetmacro{\train}{120/7}
\pgfmathsetmacro{\valid}{60/7}
\pgfmathsetmacro{\test}{30/7}
\pgfmathsetmacro{\step}{\test}
\pgfmathsetmacro{\xscale}{0.375}
\pgfmathsetmacro{\rowgap}{1.05}
\pgfmathsetmacro{\bh}{0.6}
\foreach \w in {0,1,2}{%
    \pgfmathsetmacro{\yrow}{\w * (-\rowgap)}
    \pgfmathsetmacro{\xstart}{\w * \step * \xscale}
    \pgfmathsetmacro{\xtrain}{\xstart + \train * \xscale}
    \pgfmathsetmacro{\xvalid}{\xtrain + \valid * \xscale}
    \pgfmathsetmacro{\xtest}{\xvalid + \test * \xscale}
    \pgfmathsetmacro{\ymid}{\yrow + \bh/2}
    \pgfmathsetmacro{\ybar}{\yrow + \bh}
    \pgfmathtruncatemacro{\wlabel}{\w + 1}
    \fill[traincolor!80, rounded corners=1.5pt]
        (\xstart cm, \yrow cm) rectangle (\xtrain cm, \ybar cm);
    \pgfmathsetmacro{\trainmid}{(\xstart+\xtrain)/2}
    \node[lbl] at (\trainmid cm, \ymid cm) {Training};
    \fill[validcolor!85, rounded corners=1.5pt]
        (\xtrain cm, \yrow cm) rectangle (\xvalid cm, \ybar cm);
    \pgfmathsetmacro{\valmid}{(\xtrain+\xvalid)/2}
    \node[lbl] at (\valmid cm, \ymid cm) {Validation};
    \fill[testcolor!85, rounded corners=1.5pt]
        (\xvalid cm, \yrow cm) rectangle (\xtest cm, \ybar cm);
    \pgfmathsetmacro{\testmid}{(\xvalid+\xtest)/2}
    \node[lbl] at (\testmid cm, \ymid cm) {Testing};
    \node[rowlbl] at (-0.15cm, \ymid cm) {$w\wlabel$};
}
\pgfmathsetmacro{\braceW}{0}
\pgfmathsetmacro{\braceTrain}{\train*\xscale}
\pgfmathsetmacro{\braceValid}{\train*\xscale + \valid*\xscale}
\pgfmathsetmacro{\braceTest}{\train*\xscale + \valid*\xscale + \test*\xscale}
\pgfmathsetmacro{\bracey}{\bh + 0.18}
\draw[decorate, decoration={brace, amplitude=6pt}, line width=1.2pt, traincolor!80]
    (\braceW cm, \bracey cm) -- (\braceTrain cm, \bracey cm)
    node[midway, above=7pt, font=\fontsize{10}{14}\sffamily\bfseries, text=traincolor!80]
    {$t \in \mathcal{T}^{\rm train}$};
\draw[decorate, decoration={brace, amplitude=6pt}, line width=1.2pt, validcolor!85]
    (\braceTrain cm, \bracey cm) -- (\braceValid cm, \bracey cm)
    node[midway, above=7pt, font=\fontsize{10}{14}\sffamily\bfseries, text=validcolor!85]
    {$\kappa \in \mathcal{T}^{\rm val}$};
\draw[decorate, decoration={brace, amplitude=6pt}, line width=1.2pt, testcolor!85]
    (\braceValid cm, \bracey cm) -- (\braceTest cm, \bracey cm)
    node[midway, above=7pt, font=\fontsize{10}{14}\sffamily\bfseries, text=testcolor!85]
    {$\tau \in \mathcal{T}^{\rm test}$};
\foreach \w in {0,1}{%
    \pgfmathsetmacro{\xarrow}{(\w * \step + \train + \valid + \test) * \xscale}
    \pgfmathsetmacro{\ytop}{\w * (-\rowgap) - 0.05}
    \pgfmathsetmacro{\ybot}{(\w+1) * (-\rowgap) + \bh + 0.05}
    \draw[->, gray!50, thin]
        (\xarrow cm, \ytop cm) -- (\xarrow cm, \ybot cm);
}
\pgfmathsetmacro{\axisy}{-(2*\rowgap) - \bh +0.2}
\pgfmathsetmacro{\axisend}{(120+60+30+2*\test*7)/7*\xscale + 0.4}
\draw[->, gray!60, semithick] (-0.1cm, \axisy cm) -- (\axisend cm, \axisy cm);
\node[axislbl, anchor=west] at (\axisend cm, \axisy cm) {Time};
\end{tikzpicture}
}
{Rolling window procedure for training, validation, and testing. \label{fig:rolling_windows}}
{Training, validation and testing sets for each window ($w1$, $w2$, ...). The validation set is used for tuning the hyperparameters of the classification model and the confidence thresholds for the predict-then-contextual-optimize model. Final model training is performed on the combined training and validation sets. Note that this schematic only visualizes 3 windows, while the case study in Section~\ref{sec:Results} considers 22 windows.}
\end{figure}
\begin{itemize}
    \item \textbf{Training set} $\mathcal{T}^{\rm train}(w)$: a set of training periods used to train the classifier and two linear decision policies for opportunistic long and short bids. The training data provide the historical observations of the features $\mathbf{x}_t$, e.g.,\ renewable generation and demand forecasts, and the realized price spread $\Delta\lambda_t$, from which the predictive and contextual optimization models are learned.

    \item \textbf{Validation set} $\mathcal{T}^{\rm val}(w)$: the subsequent periods, held out from training and used for hyperparameter tuning. The models are validated with various hyperparameter configurations on this set to select the optimal configuration. Finally, the model is trained on the combined training and validation sets, i.e., $\mathcal{T}^{\rm train}(w) \cup \mathcal{T}^{\rm val}(w)$.

    \item \textbf{Test set} $\mathcal{T}^{\rm test}(w)$: the final trading periods, on which out-of-sample performance is evaluated using the retrained model. No feedback from the test set is used to adjust any model component, as this would bias the training process.
\end{itemize}

After evaluating window $w$, the entire window is shifted forward by the length of the testing set and the procedure repeats for window $w+1$, yielding a sequence of non-overlapping test periods that together span the full evaluation horizon. This design ensures that all reported test results are genuinely out-of-sample, that the model is regularly retrained to adapt to distributional drifts in market conditions, and that hyperparameters are never selected using test-period data.

\section{Training and Validation}\label{sec:train_valid}

In this section, we describe the training phase of our model framework. The framework is divided into two consecutive modeling steps that are coupled during training. Step~1 is a probabilistic classification model based on two confidence thresholds and is described in Section~\ref{sec:classif}. Step~2 is a contextual optimization problem that trains a linear decision policy for each class and is described in Section~\ref{sec:optimization}.

\subsection{Classification}\label{sec:classif}
\begin{figure}
     \FIGURE
{
\begin{tikzpicture}[
node distance=3cm,
every node/.style={outer sep=0, font=\fontsize{10}{12}\selectfont},
box/.style={draw, thick, rectangle, fill=green!20,
            minimum width=3cm, minimum height=1.2cm, align=center,
            font=\fontsize{10}{12}\selectfont},
boxprofit/.style={draw, thick, rectangle, fill=orange!20,
            minimum width=2cm, minimum height=1cm, align=center,
            font=\fontsize{10}{12}\selectfont},
arrow/.style={->, thick},
feedback/.style={->, thick, dashed},
profit_feedback/.style={->, thick, dashed, orange}
]
\node (input) {$(\mathbf{x}_t,y_t)$};
\node[box, right=2cm of input] (clf)
{Classifier training\\ [2pt]$f_{\bm{\theta}}(\mathbf{x}_t, \bm{\Gamma})$};
\node[box, right=of clf] (thresh)
{Threshold function\\ [2pt]$n_t(f_{\bm{\theta}}(\mathbf{x}_t, \bm{\Gamma}),\underline{\alpha},\overline{\alpha})$};
\node[boxprofit, below=1cm of thresh] (profitopt)
{Profit calculation\\[2pt]$\underset{\underline{\alpha},\overline{\alpha}}{\max}\ \rho[\Pi_{\kappa}(\underline{\alpha},\overline{\alpha})]$};
\node[boxprofit, below=1cm of clf] (accopt)
{ROC-AUC in validation\\[2pt]$\underset{\bm{\Gamma}}{\max}\ \mathrm{AUC}(y_\kappa, f_{\bm{\theta}}(\mathbf{x}_\kappa,\bm{\Gamma}))$};
\coordinate (outend) at ($(thresh.east) + (2,0)$);
\path (clf) -- coordinate[midway] (midclf) (thresh);
\path (thresh.east) -- coordinate[midway] (midthresh) (outend);
\draw[arrow] (input) -- (clf);
\draw[arrow] (clf) -- node[above] {$[0,1]$} (thresh);
\draw[arrow] (thresh.east) -- node[above] {$\{-1,0,1\}$} (outend);
\coordinate (midthresh-down) at (midthresh |- profitopt.east);
\draw[profit_feedback] (midthresh) -- (midthresh-down) -- (profitopt.east);
\coordinate (midclf-down) at (midclf |- accopt.east);
\draw[profit_feedback] (midclf) -- (midclf-down) -- (accopt.east);
\draw[profit_feedback] (profitopt.north) -- node[right] {$\underline{\alpha},\overline{\alpha}$} (thresh.south);
\draw[profit_feedback] (accopt.north) -- node[right] {$\bm{\Gamma}$} (clf.south);
\node (input2) at (input |- accopt) {$(\mathbf{x}_\kappa, y_\kappa)$};
\draw[arrow] (input2) -- (accopt);
\end{tikzpicture}
}
{Training for Step 1: Probabilistic classification. \label{train classif}}
{The two components of the probabilistic classification (Step 1) are shown in the green boxes. The tuning of the hyperparameters is shown through the orange dashed feedback loops and the orange boxes. We tune the hyperparameters of the classifier, $\bm{\Gamma}$, based on the ROC-AUC of predictions on the validation set. The confidence thresholds $\underline{\alpha}$ and $\overline{\alpha}$ are tuned based on the downstream profit function $\rho$, which can be either the expected profit or the CVaR of profit.}
\end{figure}

In the first model step, a probabilistic binary classifier is trained to predict the sign of the price spread $\Delta\lambda_t = \lambda^{\rm DA}_t - \lambda^{\rm B}_t$. As established in Section~\ref{sec:overview}, the sign of $\Delta\lambda_t$ is mainly driving the direction of the arbitrage decision. The training procedure is illustrated in Figure~\ref{train classif}.

The probabilistic classification is performed based on two components (green boxes in Figure~\ref{train classif}). The first component (left green box) is a probabilistic binary classifier $f_{\bm{\theta}}(\mathbf{x}_t,\bm{\Gamma}): \mathbb{R}^d \to [0,1]$ that receives the features $\mathbf{x}_t \in \mathbb{R}^{d}$ of time period $t \in \mathcal{T^{\rm train}}$ as input and maps them to the binary target $y_t$ that takes the value $1$ if $\Delta\lambda_t > 0$, and $-1$ if $\Delta\lambda_t < 0$. 
Notation $\bm{\theta}$ denotes the classifier parameters, which are learned in the training phase by minimizing the class-weighted binary cross-entropy on $\mathcal{T}^{\rm train}$.
The classifier's own model hyperparameters that tune its accuracy and regularization are denoted by $\bm{\Gamma}$. They are tuned by maximizing a performance metric evaluated on the out-of-sample validation set, indexed by $\kappa \in \mathcal{T^{\rm val}}$ (left orange box) to find the best hyperparameter configuration of the classifier $\tilde{\bm{\Gamma}}$. In our case study, we use the area under the receiver operating characteristic curve (ROC-AUC) as the performance metric to select the best hyperparameters, but it is a design choice for the power trader.

In addition to the classifier hyperparameters, we introduce a deadband quantile $\delta$ that selects which training samples enter the classifier based on the absolute magnitude of the price spread $|\Delta\lambda_t|$. Samples with a near-zero spread, $|\Delta\lambda_t|\approx 0$, carry essentially no arbitrage value, since the profit is then almost independent of the bidding decision, while at the same time their labels are the most noise-sensitive, because an arbitrarily small perturbation flips the sign. We remove these low-value, high-noise samples from the classifier training set by introducing a central deadband around zero, i.e., by dropping the fraction $\delta$ of training samples with the smallest absolute spread,
\begin{equation}\label{eq:deadband}
    \mathcal{T}^{\rm train}_{\delta} = \bigl\{\, t \in \mathcal{T}^{\rm train} : |\Delta\lambda_t| > Q_\delta \,\bigr\},
\end{equation}
where $Q_\delta$ denotes the $\delta$-quantile of the absolute spread $|\Delta\lambda|$ over the training set ($\delta=0$ recovers the full training set). 
\begin{figure}
\FIGURE
{%
\begin{minipage}[b]{0.49\textwidth}
\centering
\resizebox{\textwidth}{!}{%
\begin{tikzpicture}[
    >=stealth,
    font=\fontsize{10}{12}\selectfont,
    rlbl/.style={align=center, font=\fontsize{9}{11}\selectfont},
]
\def\H{2.7}     
\def\s{1.25}    
\def\db{1.0}    
\def\dbb{2.0}   
\def\xmax{4.0}
\fill[red!16]
    plot[domain=-\xmax:-\db, samples=60] (\x, {\H*exp(-abs(\x)/\s)})
    -- (-\db,0) -- (-\xmax,0) -- cycle;
\fill[green!16]
    plot[domain=\db:\xmax, samples=60] (\x, {\H*exp(-abs(\x)/\s)})
    -- (\xmax,0) -- (\db,0) -- cycle;
\fill[gray!28]
    plot[domain=-\db:\db, samples=60] (\x, {\H*exp(-abs(\x)/\s)})
    -- (\db,0) -- (-\db,0) -- cycle;
\draw[thick] plot[domain=-\xmax:\xmax, samples=160] (\x, {\H*exp(-abs(\x)/\s)});
\draw[->, semithick] (-\xmax-0.35,0) -- (\xmax+0.5,0) node[right] {$\Delta\lambda_t$};
\draw[dashed, gray] (0,0) -- (0,\H+0.2);
\draw[thick, gray!60!black] (-\db,0) -- (-\db,{\H*exp(-\db/\s)});
\draw[thick, gray!60!black] (\db,0)  -- (\db,{\H*exp(-\db/\s)});
\draw[dashed, gray!60!black] (-\dbb,0) -- (-\dbb,{\H*exp(-\dbb/\s)});
\draw[dashed, gray!60!black] (\dbb,0)  -- (\dbb,{\H*exp(-\dbb/\s)});
\draw[<->, gray!60!black] (\db,-0.45) -- node[below] {Larger $\delta$} (\dbb,-0.45);
\node[rlbl, red!55!black]   at (-2.45,0.9) {Negative spread\\[-2pt]$y_t{=}-1$};
\node[rlbl, green!45!black] at ( 2.45,0.9) {Positive spread\\[-2pt]$y_t{=}1$};
\node[rlbl] at (0,\H+0.7) {Drop lowest $\delta$ fraction};
\end{tikzpicture}%
}
\subcaption{Deadband quantile $\delta$.\label{fig:deadband}}
\end{minipage}
\hfill
\begin{minipage}[b]{0.49\textwidth}
\centering
\resizebox{\textwidth}{!}{%
\begin{tikzpicture}[
    >=stealth,
    font=\fontsize{10}{12}\selectfont,
    rlbl/.style={align=center, font=\fontsize{9}{11}\selectfont},
]
\def\W{6}        
\def\Hh{3.4}     
\def\azero{0.3}  
\def\aone{0.7}   
\fill[red!16]   (0,0)            rectangle (\W,{\azero*\Hh});  
\fill[gray!16]  (0,{\azero*\Hh}) rectangle (\W,{\aone*\Hh});   
\fill[green!16] (0,{\aone*\Hh})  rectangle (\W,\Hh);           
\draw[->, semithick] (0,0) -- (\W+0.45,0) node[right] {Context $x_t$};
\draw[->, semithick] (0,0) -- (0,\Hh+0.45) node[above] {$f_{\tilde{\bm{\theta}}}(x_t, \tilde{\bm{\Gamma}})$};
\foreach \p in {0,0.5,1}{
    \draw (0,{\p*\Hh}) -- (-0.1,{\p*\Hh}) node[left] {\p};
}
\draw[dashed, thick] (0,{\aone*\Hh})  -- (\W,{\aone*\Hh})  node[right] {$\overline{\alpha}$};
\draw[dashed, thick] (0,{\azero*\Hh}) -- (\W,{\azero*\Hh}) node[right] {$\underline{\alpha}$};
\draw[very thick, blue!70!black]
    (0,{0.10*\Hh}) -- (0.7,{0.10*\Hh}) -- (0.7,{0.20*\Hh}) --
    (1.5,{0.20*\Hh}) -- (1.5,{0.42*\Hh}) -- (2.2,{0.42*\Hh}) --
    (2.2,{0.80*\Hh}) -- (3.1,{0.80*\Hh}) -- (3.1,{0.55*\Hh}) --
    (3.8,{0.55*\Hh}) -- (3.8,{0.86*\Hh}) -- (4.7,{0.86*\Hh}) --
    (4.7,{0.93*\Hh}) -- (\W,{0.93*\Hh});
\foreach \px/\py in {0.35/0.10, 1.1/0.20}
    \fill[red!70!black] (\px,{\py*\Hh}) circle (1.4pt);
\foreach \px/\py in {1.85/0.42, 3.45/0.55}
    \fill[gray!55!black] (\px,{\py*\Hh}) circle (1.4pt);
\foreach \px/\py in {2.65/0.80, 4.25/0.86, 5.35/0.93}
    \fill[green!55!black] (\px,{\py*\Hh}) circle (1.4pt);
\node[rlbl, green!45!black] at (1.55,{0.88*\Hh}) {Opportunistic long bid\\[-2pt]$n_t{=}{+}1$};
\node[rlbl, black!60]       at (1,{0.50*\Hh}) {Arbitrage-free\\[-2pt]$n_t{=}0$};
\node[rlbl, red!55!black]   at (4,{0.13*\Hh}) {Opportunistic short bid\\[-2pt]$n_t{=}{-}1$};
\end{tikzpicture}%
}
\subcaption{Confidence thresholds $\underline{\alpha},\overline{\alpha}$.\label{fig:alpha_thresholds}}
\end{minipage}%
}
{Schematics of the classifier hyperparameters.\label{fig:hparam_schematics}}
{(a) Schematic distribution of the realized price spread $\Delta\lambda_t$ over a training set. The classifier is trained only on samples outside the central deadband (gray). The fraction $\delta$ of samples with the smallest $|\Delta\lambda_t|$ is dropped, keeping the clearer negative spread ($y_t{=}-1$) and positive spread ($y_t{=}1$) tails. (b) Simplified schematic of the classifier output $f_{\bm{\theta}}(x_t,\tilde{\bm{\Gamma}})$ for a one-dimensional feature $x_t$. The thresholds $\underline{\alpha}$ and $\overline{\alpha}$ are two horizontal cuts. Above $\overline{\alpha}$ the sample is classified to make an opportunistic long bid ($n_t{=}{+}1$, green), below $\underline{\alpha}$ to make an opportunistic short bid ($n_t{=}{-}1$, red), and within the band $[\underline{\alpha},\overline{\alpha}]$ to make an arbitrage-free bid ($n_t{=}0$, gray). For a tree-based model, $f_{\tilde{\bm{\theta}}}(x_t, \tilde{\bm{\Gamma}})$ is a step function.}
\end{figure}
Figure~\ref{fig:deadband} illustrates in a schematic how the training set shrinks as $\delta$ grows. The deadband quantile $\delta$ is also tuned to maximize the ROC-AUC on the validation set, i.e., it is searched over a common grid with $\bm{\Gamma}$.

Using the trained classifier with the best hyperparameter configuration $\tilde{\bm{\Gamma}}$ and $\tilde{\delta}$, we introduce two confidence thresholds $\underline{\alpha} \in [0, 0.5]$ and $\overline{\alpha} \in [0.5, 1]$ as the second component (right green box in Figure~\ref{train classif}) to make the final probabilistic classification. This allows us to control the confidence of the final prediction separately for long and short bids. The predicted class $n_t \in \{-1, 0, 1\}$ is then determined by
\begin{equation}\label{eq:threshold_rule}
    n_t = \begin{cases}
        \phantom{-}1  & \text{if } f_{\bm{\tilde{\theta}}}(\mathbf{x}_t,\tilde{\bm{\Gamma}}) \geq \overline{\alpha}, \\
        -1            & \text{if } f_{\bm{\tilde{\theta}}}(\mathbf{x}_t,\tilde{\bm{\Gamma}}) \leq \underline{\alpha}, \\
        \phantom{-}0  & \text{otherwise,}
    \end{cases}
\end{equation}
where $n_t = 1$ corresponds to making an opportunistic long bid (e.g.,\ bid maximum in the day-ahead market), $n_t = -1$ to making an opportunistic short bid (e.g.,\ bid minimum), and $n_t = 0$ to making an arbitrage-free bid. Geometrically, as illustrated in a simplified schematic for a single feature $x_t$ (one-dimensional) in Figure~\ref{fig:alpha_thresholds}, the thresholds are two horizontal cuts through the learned score function. The thresholds $\underline{\alpha}$ and $\overline{\alpha}$ tune the trade-off between arbitrage exploitation and confidence and can be asymmetric, so the trader can demand different confidence levels for opportunistic long and short bids. When $\underline{\alpha} = \overline{\alpha} = 0.5$, every sample is assigned an opportunistic bid, i.e., there are no arbitrage-free bids. As $\underline{\alpha}$ decreases toward $0$ and $\overline{\alpha}$ increases toward $1$, the number of arbitrage-free bids increases, reducing the exploitation rate of opportunistic bids but improving their confidence.
The threshold rule~(\ref{eq:threshold_rule}) is a reject option in the sense of~\citet{Chow1970} with asymmetric costs. Since the classifier output is not calibrated (see the end of this section), this reading is qualitative: a higher $\overline{\alpha}$ expresses that a misdirected long bid is costlier, relative to abstaining, than a misdirected short bid.

Since the values of $\underline{\alpha}$ and $\overline{\alpha}$ determine which samples enter the policy optimization in Step~2, they are not tuned to minimize the classification loss but value-oriented instead, i.e., to maximize a measure $\rho$ of the profit, which could be the expectation or CVaR measure, on the validation set (right orange box in Figure~\ref{train classif}):
\begin{equation}
    \underset{\underline{\alpha}, \overline{\alpha}}{\max}\ \rho\!\left[\Pi_\kappa(\underline{\alpha}, \overline{\alpha})\right],
\end{equation}
where $\Pi_\kappa(\underline{\alpha}, \overline{\alpha})$ is the realized profit on the validation set, $\kappa \in \mathcal{T^{\rm val}}$, for the given confidence thresholds $\underline{\alpha}$ and $\overline{\alpha}$. It is calculated using the optimization model from Section~\ref{sec:optimization} that is trained on the training set and then yields the out-of-sample bidding decisions for the validation set. Using these bidding decisions, we can calculate the final profit of the validation set, as described in Section~\ref{sec:real_time}. Depending on the risk preference of the decision maker, $\rho$ is either the expectation  $\mathbb{E}[\Pi_\kappa(\underline{\alpha}, \overline{\alpha})]$ or the $\text{CVaR}[\Pi_\kappa(\underline{\alpha}, \overline{\alpha})]$ of profit, which penalizes large losses in the tail of the profit distribution.

Since Step~2 only receives the class $n_t$ and the thresholds $(\underline{\alpha},\overline{\alpha})$ are tuned afterwards on the validation profit, the bidding decisions depend on the classifier output $f_{\bm{\theta}}(\mathbf{x}_t,\tilde{\bm{\Gamma}})$ only through the ranking of the samples, not through its absolute values. Any strictly increasing transformation of the output, e.g., Platt scaling, combined with correspondingly transformed thresholds, yields the same classes $n_t$. The relevant property of the classifier is therefore how well it ranks samples with a positive spread above samples with a negative spread, which is what the ROC-AUC measures. We use the ROC-AUC to select $\tilde{\bm{\Gamma}}$ and $\tilde{\delta}$, while the economic consequences of a wrong direction, which depend on the realized price spread, the wind power production and the policies of Step~2, enter only through the profit-based tuning of the thresholds. Hence, the classifier output does not need to be a calibrated probability. The thresholds $(\underline{\alpha},\overline{\alpha})$ should therefore be read as the strength of evidence required before an opportunistic bid is placed, not as the probability that the predicted direction is correct. This is sufficient as the thresholds are tuned on the validation profit and not set a priori.
\subsection{Optimization}\label{sec:optimization}

Given the predicted class $n_t \in \{-1, 0, 1\}$ from Step~1, we introduce a contextual optimization model (\textit{Policies}) as Step~2 that translates the prediction into a day-ahead quantity bid $p_t^{\rm DA}$ to maximize the profit. The model can be configured for different assets. We focus here on a standalone wind farm (wind-only) and an HPP consisting of a co-located wind farm and electrolyzer.

Similar to \citet{Heiser2025}, we learn a linear decision policy $\pi(n_t,x_t)$ that maps the features $\mathbf{x}_t$ to the bidding decision, instead of fixing the bid at a boundary value. However, unlike~\citet{Heiser2025}, we learn two separate policies, one for the long and one for the short class, i.e., for $n_t=1$ and $n_t=-1$. In our case, the optimal bidding decision $p_t^{\rm DA}$  is parameterized as
\begin{equation}\label{eq:policy}
    p_t^{\rm DA} = \hat{P}_t^{\rm W} + z_t, \qquad z_t = \pi(n_t,\mathbf{x}_t) = \mathbf{q}_{n_t}\, \mathbf{x}_t^\top,
\end{equation}
where $\mathbf{q}_{n_t} \in \mathbb{R}^{|\mathbf{x_t}|}$ is a class-specific parameter vector. The two vectors, $\mathbf{q}_{-1}$ and $\mathbf{q}_1$, are learned simultaneously by solving the optimization program shown below over the training set $\mathcal{T}^{\rm train}$. For the HPP case it reads as follows:
\begin{subequations}\label{eq:5}
\allowdisplaybreaks[0]
\begin{breakablealign}
\max_{\substack{z_t,\, p_t^{\rm DA},\, p_t^{\rm B},\, p_t^{\rm H},\\ h_t,\, \mathbf{q}_{n_t},\, \xi_t,\, \zeta}}
\quad &
\frac{\beta}{T}
\sum_{d \in \mathcal{T}^{\rm train}}
\sum_{t \in \mathcal{T}^{\rm train}(d)}
\left(
p_t^{\rm DA}{\lambda}_t^{\rm DA}
+
p_t^{\rm B}{\lambda}_t^{\rm B}
+
h_t{\lambda}^{\rm H}
\right) \nonumber \\
& \quad +
(1-\beta)
\left(
\zeta
-
\frac{1}{T\epsilon}
\sum_{d \in \mathcal{T}^{\rm train}}
\sum_{t \in \mathcal{T}^{\rm train}(d)}
\xi_t
\right)
\label{eq:Optimization_a} \\
\text{s.t.}\quad
& (\ref{eq:policy})
&& \forall d,t \nonumber \\
& P_t^{\rm W}
= p_t^{\rm DA} + p_t^{\rm B} + p_t^{\rm H}
&& \forall d,t
\label{eq:Optimization_b} \\
& -\overline{P}^{\rm H}
\le p_t^{\rm DA}
\le \overline{P}^{\rm W}
&& \forall d,t
\label{eq:Optimization_c} \\
& \underline{P}^{\rm H}
\le p_t^{\rm H}
\le \overline{P}^{\rm H}
&& \forall d,t
\label{eq:Optimization_d} \\
& h_t \le A_s p_t^{\rm H} + B_s
&& \forall s,d,t
\label{eq:Optimization_e} \\
& \sum_{t \in \mathcal{T}^{\rm train}(d)} h_t
\ge \underline{H}
&& \forall d
\label{eq:Optimization_f} \\
& z_t \le 0
&& \forall d,t : n_t = -1
\label{eq:Optimization_i} \\
& z_t \ge 0
&& \forall d,t : n_t = 1
\label{eq:Optimization_j} \\
& z_t = 0
&& \forall d,t : n_t = 0
\label{eq:Optimization_k} \\
& \xi_t \ge
\zeta
-
\left(
p_t^{\rm DA}{\lambda}_t^{\rm DA}
+
p_t^{\rm B}{\lambda}_t^{\rm B}
+
h_t\lambda^{\rm H}
\right)
&& \forall d,t
\label{eq:Optimization_l} \\
& \xi_t \ge 0 && \forall d,t. \label{eq:Optimization_n}
\end{breakablealign}
\end{subequations}

The optimization program~\eqref{eq:5} is a linear program that learns the class-specific policy parameters $\mathbf{q}_{n_t}$ across the full training period. The objective~\eqref{eq:Optimization_a} maximizes a weighted combination of the mean profit and the CVaR$_\epsilon$ of the profit, following the Rockafellar-Uryasev formulation \citep{Rockafellar2000}, for all days $d$ in the training period. The first term averages the profit from the day-ahead market, the balancing market, and hydrogen sales across all training periods, using the realized prices ${\lambda}_t^{\rm DA}$, ${\lambda}_t^{\rm B}$, and ${\lambda}^{\rm H}$. The second term, $\zeta - \frac{1}{T\epsilon}\sum_{d,t}\xi_t$, is the CVaR$_\epsilon$ of the profit, i.e., the expected profit in the worst $\epsilon$ fraction of training periods, where $\epsilon \in (0,1]$ is the tail probability and $T$ is the total number of samples of the training period. The scalar $\beta \in [0,1]$ is the weight on the objective terms. Pure expected-profit maximization is achieved by $\beta = 1$, while smaller values of $\beta$ shift the weight toward the tail risk of the profit distribution. Constraint~\eqref{eq:Optimization_b} is the power balance between the realized wind production $P_t^{\rm W}$, the day-ahead bid $p_t^{\rm DA}$, the balancing position $p_t^{\rm B}$, and the electrolyzer consumption $p_t^{\rm H}$. Constraint~\eqref{eq:Optimization_c} bounds the day-ahead bid. The upper bound $\overline{P}^{\rm W}$ represents the maximum available wind capacity, while the lower bound $-\overline{P}^{\rm H}$ allows the HPP to procure electricity from the grid via the day-ahead market to feed the electrolyzer when it is economically favorable. Constraint~\eqref{eq:Optimization_d} restricts the power consumption of the electrolyzer to its operational range $[\underline{P}^{\rm H}, \overline{P}^{\rm H}]$, where $\underline{P}^{\rm H}$ is the minimum stable load required by the device. Constraint~\eqref{eq:Optimization_e} represents a piecewise linear approximation of the non-linear production curve of the electrolyzer following~\citet{Raheli2023AScheduling}. The produced hydrogen $h_t$ is upper-bounded by linear segments $s$ with slopes $A_s$ and intercepts $B_s$. Constraint~\eqref{eq:Optimization_f} imposes a minimum daily hydrogen production $\underline{H}$ aggregated over all delivery periods of each day $d$, reflecting a contractual offtake obligation or operational target. As given by~\eqref{eq:policy}, the day-ahead bid is parameterized as the wind power forecast $\hat{P}_t^{\rm W}$ plus a signed deviation $z_t$, where $z_t$ is the output of the class-specific linear policy, i.e., the inner product of the parameter vector $\mathbf{q}_{n_t}$ and the transposed feature vector $\mathbf{x}_t^\top$. Constraints~\eqref{eq:Optimization_i}--\eqref{eq:Optimization_k} enforce the directional intent of the classifier for each class: $z_t \leq 0$ for opportunistic short bids ($n_t = -1$), $z_t \geq 0$ for opportunistic long bids ($n_t = 1$) and $z_t = 0$ for arbitrage-free bids ($n_t = 0$), i.e., bidding exactly the wind power forecast. Together, these three constraints ensure that the learned policies on the training set never contradict the classification from Step 1. Finally,~\eqref{eq:Optimization_l}--\eqref{eq:Optimization_n} implement the CVaR$_\epsilon$ auxiliary variables following \citet{Rockafellar2000}. The auxiliary variable $\xi_t$ in~\eqref{eq:Optimization_l} captures the shortfall of the profit below the value-at-risk level $\zeta$. Constraint~\eqref{eq:Optimization_n} enforces the non-negativity of $\xi_t$, ensuring that only downside deviations contribute to the tail penalty. The value-at-risk level $\zeta$ is a free variable optimized jointly with the policy parameters, and the CVaR expression in the objective is concave in $\zeta$, preserving the linearity of the overall program.

The decisions in~(\ref{eq:5}) fall into two groups. The day-ahead quantity bid $p_t^{\rm DA}$ must be submitted before the day-ahead gate closure. Through~(\ref{eq:policy}), it depends only on the forecast $\hat{P}^{\rm W}_t$, the class $n_t$, and the features $\mathbf{x}_t$, which contain only information available at the time of decision making (Appendix~\ref{app:features}). Hence, the policy parameters $\mathbf{q}_{-1}$ and $\mathbf{q}_1$ are learned with realized parameters in~(\ref{eq:5}) but can be applied directly  to $\mathbf{x}_\tau$ in testing without any future information. The variables $p_t^{\rm B}$, $p_t^{\rm H}$ and $h_t$ are real-time decisions made after the day-ahead market clearing and close to delivery. In the policy training~(\ref{eq:5}), they are optimized with the realizations $P_t^{\rm W}$ and $\lambda_t^{\rm B}$, while in testing the dispatch follows~(\ref{eq:hpp_balancing}), which will be presented later.

For the wind-only case, the contextual optimization problem simplifies in four ways: (i) the hydrogen revenue term $h_t{\lambda}^{\rm H}$ is dropped from both the profit sums in the objective function~\eqref{eq:Optimization_a} and the CVaR auxiliary constraint~\eqref{eq:Optimization_l}, (ii) the power balance~\eqref{eq:Optimization_b} reduces to $P_t^{\rm W} = p_t^{\rm DA} + p_t^{\rm B}$, (iii) the lower bound on the day-ahead bid in~\eqref{eq:Optimization_c} tightens from $-\overline{P}^{\rm H}$ to $0$, since there is no electrolyzer to consume grid power, and (iv) the electrolyzer constraints~\eqref{eq:Optimization_d}--\eqref{eq:Optimization_f} are removed entirely. All remaining constraints are identical to the HPP formulation.

Note that Step~1 and Step~2 are trained with different objectives. Step~1 minimizes a classification loss on the sign of $\Delta\lambda_t$, whereas Step~2 maximizes the weighted combination of mean profit and CVaR in~(\ref{eq:Optimization_a}). Without~(\ref{eq:Optimization_i}) and~(\ref{eq:Optimization_j}), the two steps could disagree on the direction of the opportunistic bid. We let Step~1 decide the direction, as the trader should only deviate from the arbitrage-free bid if the system status can be predicted with sufficient confidence, and only in the predicted direction. For samples in which both steps agree,~(\ref{eq:Optimization_i}) and~(\ref{eq:Optimization_j}) are not binding and Step~2 chooses the extent. In case of disagreement, the constraints bind and shrink the deviation toward zero, i.e., toward the arbitrage-free bid, instead of reversing the direction. In addition, the classifier is a non-linear model that can capture more complex relations between the features and the sign of the price spread than a linear policy. Letting Step~2 decide the direction would require embedding the classifier into problem~(\ref{eq:5}), potentially rendering it computationally intractable. Finally, the two steps are not fully decoupled, since the confidence thresholds are tuned on the validation profit obtained with the policies of Step~2 (Section~\ref{sec:classif}). Future work could aim to fully close the loop between both steps via backpropagation of the downstream profit through problem~(\ref{eq:5}) or a change of the objective of the classifier to a decision-focused loss~\citep{Elmachtoub2020SmartOptimize}.

\section{Testing}\label{sec:testing}
\begin{figure}
     \FIGURE
 {
 \resizebox{\textwidth}{!}{
\begin{tikzpicture}[
    x=0.75cm, y=1cm,
    node distance=1cm and 2.8cm,
    boxA/.style={draw, thick, rectangle,
                 fill=blue!20, minimum width=3cm, minimum height=1.2cm,
                 align=center, font=\fontsize{9}{12}\selectfont},
    boxB/.style={draw, thick, rectangle,
                 fill=green!20, minimum width=2cm, minimum height=1.2cm,
                 align=center, font=\fontsize{9}{12}\selectfont},
    boxC/.style={draw, thick, rectangle,
                 fill=orange!20, minimum width=3cm, minimum height=1.2cm,
                 align=center, font=\fontsize{9}{12}\selectfont},
    group/.style={draw, thick, dashed, inner sep=12pt},
    arrow/.style={->, thick},
    every node/.style={font=\fontsize{9}{12}\selectfont}
]
\node[boxB] (class)
{Classifier\\ $f_{\tilde{\bm{\theta}}}(\mathbf{x}_\tau, \tilde{\bm{\Gamma}})$};
\node[boxB, right=0.8cm of class] (thresh)
{Threshold \\ $n_\tau(\tilde{\underline{\alpha}},\tilde{\overline{\alpha}},f_{\tilde{\bm{\theta}}}(\mathbf{x}_\tau, \tilde{\bm{\Gamma}}))$};
\node[boxA, right=1.7cm of thresh] (opt)
{Policy\\ $\pi_{\tilde{q}}(\mathbf{x}_\tau,n_\tau(\tilde{\underline{\alpha}},\tilde{\overline{\alpha}},f_{\tilde{\bm{\theta}}}(\mathbf{x}_\tau, \tilde{\bm{\Gamma}}))$};
\node (context) at ($(class)!0.5!(opt) + (0,2cm)$)
{$\mathbf{x}_\tau$};
\node[above=0.7cm of thresh] (alpha) {$\tilde{\underline{\alpha}},\tilde{\overline{\alpha}}$};
\node[boxC, right=1.7cm    of opt] (profit)
{Ex-post profit opt.\\ $\max g(p^{\rm DA}_\tau(\cdot),\lambda^{\rm DA}_\tau,\ \lambda^{\rm B}_\tau)$\\ $\text{s.t. } h(p^{\rm DA}_\tau(\cdot)) \leq 0$};
\node (realized) at ($(profit) + (0,1.7cm)$)
{$\lambda^{\rm DA}_\tau,\ \lambda^{\rm B}_\tau$};
\draw[arrow] (context) -| (class);
\draw[arrow] (context) -| (opt);
\draw[arrow] (class) -- node[above] {$[0,1]$} (thresh);
\draw[arrow] (alpha) -- (thresh);
\draw[arrow] (thresh) -- node[above, fill=white, inner sep=2pt] {$\{-1,0,1\}$} (opt);
\draw[arrow] (opt) -- node[above, fill=white, inner sep=2pt] {$p^{\rm DA}_\tau(\lambda^{\rm DA})$} (profit);
\draw[arrow] (realized) -- (profit);
\begin{scope}[on background layer]
\node[group, fit=(class)(thresh), label={[anchor=north, yshift=-20pt, font=\fontsize{10}{12}\selectfont]south:Step 1: When and what direction?}] (g1) {};
\node[group, fit=(opt), label={[anchor=north, yshift=-20pt, font=\fontsize{10}{12}\selectfont]south:Step 2: What extent?}] (g2) {};
\node[group, fit=(profit), label={[anchor=north, yshift=-17pt, font=\fontsize{10}{12}\selectfont]south:Step 3: Ex-post calculation}] (g3) {};
\node[below=0.7cm of profit] (output) {Profit};
\draw[arrow] (profit) -- (output);
\end{scope}
\end{tikzpicture}
}
}
{Full testing phase. \label{fig:testing}}
{All steps of the testing phase are shown in the three dashed boxes. The box colors refer to the decision framework in Figure~\ref{fig:decision framework}. For a given context $\mathbf{x}_\tau$ of the testing set $\tau \in \mathcal{T}^{\rm test}$, the trained classifier and tuned thresholds (green boxes) classify the sample in Step 1, deciding whether to make an opportunistic long (1) or short bid (-1) or an arbitrage-free bid (0). In Step 2, the trained policy of the class yields the day-ahead bid as a function of the day-ahead price, which is then used for final ex-post profit calculation in Step 3. All function parameters and hyperparameters are learned or tuned in the training phase, as described in Section~\ref{sec:train_valid}, and are symbolized by $\tilde{(\cdot)}$.}
\end{figure}

The testing phase applies the fully trained model to unseen test periods $\tau \in \mathcal{T}^{\rm test}$, as illustrated in Figure~\ref{fig:testing}. At each test period $\tau$, the contextual feature vector $\mathbf{x}_\tau$, available at gate closure of the day-ahead market, is processed sequentially through three steps. All model parameters and hyperparameters $\tilde{\mathbf{q}}$,  $\tilde{\bm{\theta}}$, $\tilde{\delta}$, $\tilde{\bm{\Gamma}}$, $\tilde{\underline{\alpha}}$, and $\tilde{\overline{\alpha}}$ are fixed at the values determined during the training phase and are not updated during the testing phase.

\paragraph{Step 1: When and what direction?}
The trained classifier $f_{\tilde{\bm{\theta}}}(\mathbf{x}_\tau, \tilde{\bm{\Gamma}})$ maps the features to a score $\hat{p}_\tau \in [0,1]$, where larger values indicate stronger evidence for $\Delta\lambda\tau > 0$. Applying the tuned confidence thresholds $\tilde{\underline{\alpha}}$ and $\tilde{\overline{\alpha}}$ as described in Section~\ref{sec:classif}, the threshold function assigns the predicted class $n_\tau \in \{-1, 0, 1\}$: $n_\tau = 1$ (opportunistic long bid), $n_\tau = -1$ (opportunistic short bid), or $n_\tau = 0$ (arbitrage-free bid).
\paragraph{Step 2: What extent?}
Given $n_\tau$, the trained linear policy $z_\tau=\pi_{\tilde{\mathbf{q}}}(n_\tau,\mathbf{x}_\tau) = \tilde{\mathbf{q}}_{n_\tau} \mathbf{x}_\tau^\top$ yields the signed deviation $z_\tau$, and the final day-ahead bid follows from $p_\tau^{\rm DA} = \hat{P}_\tau^{\rm W} + z_\tau$, as in~\eqref{eq:policy}. If the resulting bid violates the physical bounds, it is projected to the nearest feasible point.

\paragraph{Step 3: Profit evaluation.}
With $p_\tau^{\rm DA}$ committed, the realized day-ahead and balancing prices $\lambda_\tau^{\rm DA}$ and $\lambda_\tau^{\rm B}$ become available ex-post. The profit is calculated as described in Section~\ref{sec:real_time}, where the procedure differs between the wind-only and HPP cases.

\subsection{Profit Evaluation}\label{sec:real_time}

Once the day-ahead bid $p_\tau^{\rm DA}$ is committed in Step 2, the profit in Step 3 is computed differently depending on whether the asset is a standalone wind farm or an HPP. Note that we assume that the day-ahead price can be forecasted well, i.e., if prices become negative, we adjust the day-ahead bid to its minimum, i.e., $p^{\rm DA}=0$ (wind-only) or $p^{\rm DA}=-\overline{P}^{\rm H}$ (HPP), which reflects the pragmatic decision of a power trader in case of negative day-ahead prices. As this adjustment is made for all models including the benchmarks, this is equivalent to removing the samples with negative day-ahead prices.

\paragraph{Wind-only case.}
For a standalone wind farm, no further operational decision is required after the day-ahead bid is submitted. Any deviation between the realized wind production $P_\tau^{\rm W}$ and the committed bid is settled in the balancing market, so the balancing deviation and  profit follow directly as
\begin{equation}
    p_\tau^{\rm B} = P_\tau^{\rm W} - p_\tau^{\rm DA}, \qquad
    \Pi_\tau = \lambda_\tau^{\rm DA}\, p_\tau^{\rm DA} + \lambda_\tau^{\rm B}\, p_\tau^{\rm B}.
\end{equation}

\paragraph{HPP case.}
For the HPP, committing the day-ahead bid $p_\tau^{\rm DA}$ leaves a residual wind power quantity that must be allocated between the electrolyzer and the balancing market. Since the electrolyzer can be actively dispatched before physical delivery, its consumption is determined by solving a deterministic linear program using the balancing price forecast $\hat{\lambda}_\tau^{\rm B}$ and the wind power forecast $\hat{P}_\tau^{\rm W}$, both available before delivery:
\begin{subequations}\label{eq:hpp_balancing}
\begin{align}
\max_{p_\tau^{\rm B},\, p_\tau^{\rm H},\, h_\tau} \quad &
\sum_{\tau} \left( \hat{\lambda}_\tau^{\rm B}\, p_\tau^{\rm B} + \lambda^{\rm H} h_\tau \right) \label{eq:hpp_bal_obj}\\
\text{s.t.} \quad
& p_\tau^{\rm B} + p_\tau^{\rm H} = \hat{P}_\tau^{\rm W} - p_\tau^{\rm DA}
    && \forall\,\tau \label{eq:hpp_bal_pb}\\
& \underline{P}^{\rm H} \le p_\tau^{\rm H} \le \overline{P}^{\rm H}
    && \forall\,\tau \label{eq:hpp_bal_elec}\\
& h_\tau \le A_s\, p_\tau^{\rm H} + B_s
    && \forall\,s,\tau \label{eq:hpp_bal_h2}\\
& \textstyle\sum_{\tau \in \mathcal{T}(d)} h_\tau \ge \underline{H}
    && \forall\,d. \label{eq:hpp_bal_daily}
\end{align}
\end{subequations}
The objective function~\eqref{eq:hpp_bal_obj} maximizes the sum of balancing revenue and hydrogen sales revenue, using the forecasted balancing price $\hat{\lambda}_\tau^{\rm B}$ and the fixed hydrogen price $\lambda^{\rm H}$. The power balance~\eqref{eq:hpp_bal_pb} allocates the forecast residual between electrolyzer consumption $p_\tau^{\rm H}$ and balancing deviation $p_\tau^{\rm B}$. Constraints~\eqref{eq:hpp_bal_elec} and~\eqref{eq:hpp_bal_h2} enforce the electrolyzer's operational bounds and the piecewise linear hydrogen production efficiency, identical to the training problem~\eqref{eq:5}. Constraint~\eqref{eq:hpp_bal_daily} enforces the minimum daily hydrogen production obligation $\underline{H}$.

Given the optimal dispatch $(p_\tau^{\rm H*},\, h_\tau^*)$, 
the realized balancing volume is
\begin{equation}
    p_\tau^{\rm B} =  P_\tau^{\rm W} - p_\tau^{\rm DA} - p_\tau^{\rm H*},
\end{equation}
and the corresponding realized profit is
\begin{equation}
    \Pi_\tau = \lambda_\tau^{\rm DA}\, p_\tau^{\rm DA} + \lambda_\tau^{\rm B}\, p_\tau^{\rm B} + \lambda^{\rm H} h_\tau^*.
\end{equation}

Note that while the electrolyzer dispatch is optimized against the forecasted balancing price $\hat{\lambda}_\tau^{\rm B}$, the balancing revenue is settled ex-post at the realized price $\lambda_\tau^{\rm B}$. Unless stated otherwise, the results in Section~\ref{sec:Results} use $\hat{\lambda}^{\rm B}_\tau = \lambda^{\rm B}_\tau$ and $\hat{P}^{\rm W}_\tau = P^{\rm W}_\tau$, i.e., perfect foresight of the balancing price and wind power production in the electrolyzer dispatch.

\section{Numerical Results}\label{sec:Results}
This section presents a numerical analysis of the proposed framework. All source code is publicly available in~\citet{Heiser2026Code}.
\subsection{Case Studies}\label{sec:case_studies}
\paragraph{Data.}
Our case study considers two European bidding zones, DK1 (western Denmark) and DE/LU (Germany/Luxembourg), and two portfolios, a standalone wind farm (wind-only) and a wind farm co-located with an electrolyzer (HPP). Performance is evaluated via the rolling-window procedure, as previously illustrated in Figure~\ref{fig:rolling_windows}. The data considered for these studies range from April~2025 to February~2026. Since the data are available at different resolutions (60 and 15 minutes), we average the data to a 60-minute resolution. The physical asset is a real 7.2~MW wind farm ($\overline{P}^{\rm W}=7.2$~MW) with a point forecast $\hat{P}^{\rm W}$, optionally coupled with an electrolyzer of half the wind farm capacity ($\overline{P}^{\rm H}=3.6$~MW). The hydrogen price is set to €2/kg and the electrolyzer minimum stable load to 10\% of its nominal capacity. The hydrogen production efficiency curve is approximated with two linear segments, following the HYP-L method of~\citet{Raheli2023AScheduling}. A minimum daily hydrogen production requirement of 100~kg (approximately 5~MWh of electrolyzer energy consumption) is enforced.

Each rolling window uses a 7-day test period (models are retrained weekly), a 4-month training set, and a 1-month validation set, yielding 22 windows in total. A dataset consisting of 200+ features is created using publicly available data from ENTSO-E, Energinet and Open-Meteo, including lagged and forecasted data of relevant energy markets and weather. A full overview of the features in each category is given in Appendix~\ref{app:features}.
All features are z-score normalized. Feature selection is performed on each rolling window using SHAP importance values \citep{SHAP}, retaining only features with an importance above~0.6.

\paragraph{Classification.}
In the training phase of the classifier, we ignore the data with a zero  price spread, as no arbitrage is possible in these cases, i.e., the profit is independent of the day-ahead bidding decision, and they only add noise to the model. For each model, hyperparameters are selected by grid search on the validation set using the ROC-AUC as the criterion. The deadband quantile $\delta$ is tuned over $[0.0,0.8]$. The confidence thresholds are tuned over the grid $\underline{\alpha} \in \{0.15,\,0.25,\,0.35,\,0.45\}$ and $\overline{\alpha} \in \{0.55,\,0.65,\,0.75,\,0.85\}$. Among the tested classification models such as statistical, tree-based, and neural network approaches, LightGBM (LGBM) achieves the best out-of-sample classification performance and is used in all subsequent result sections. The model specifications and hyperparameter grid of the LGBM model are given in Appendix~\ref{app:models}.

\paragraph{Model Overview.}
We compare our proposed predict-then-contextual-optimize model with 6 benchmarks in the following analysis. All models are summarized in Table~\ref{tab:models}. Our proposed \textit{Classification + Policies} model pairs the Step 1 classifier of Section~\ref{sec:classif} with the Step 2 linear policy models of Section~\ref{sec:optimization}, learned via contextual optimization. 

\begin{table}[h!]
\TABLE
{Overview of the 7 models compared in the numerical results.\label{tab:models}}
{\begin{tabular}{@{}l l p{0.52\textwidth}@{}}
\hline\up
Model & Type & Description \\ \hline\up
\textit{Hindsight} & Benchmark & Bids under perfect foresight of the realized day-ahead and balancing prices. \\
\textit{Bid Forecast} & Benchmark & Bids the wind power point forecast arbitrage-free into the day-ahead market. \\
\textit{Single Policy} & Benchmark & Learns a single linear decision policy, without the classification step, proposed in~\citet{Heiser2025}. \\
\textit{Classification + Hindsight} & Benchmark & Classification as in Section~\ref{sec:classif}, paired with a bid set under perfect price foresight within the predicted class. \\
\textit{Classification + All-or-Nothing} & Benchmark & Classification as in Section~\ref{sec:classif}, paired with the binary all-or-nothing bidding rule. \\
\textit{Hindsight + Policies} & Benchmark & Oracle classification, paired with the learned linear policies per class as in Section~\ref{sec:optimization}. \\
\textit{Classification + Policies} & Proposed & Classification as in Section~\ref{sec:classif}, paired with the learned linear policies per class as in Section~\ref{sec:optimization}.\down \\ \hline
\end{tabular}}{}
\end{table}

\subsection{Profit Over All Testing Windows}\label{sec:results_profit}
We start by calculating the profit improvement of our proposed \textit{Classification + Policies} model in comparison to the simple arbitrage-free bidding strategy of the \textit{Bid Forecast} model that always bids the forecasted wind power production into the day-ahead market. 
For all models, we assume perfect information about the realized wind power production and balancing price in the real-time dispatch of the electrolyzer in~(\ref{eq:hpp_balancing}). A sensitivity analysis on the balancing price prediction in real time is provided in Section~\ref{sec:results_balancing_sensitivity}.
\begin{figure}
    \FIGURE
    {%
        \begin{minipage}[t]{\textwidth}
            \centering
            \input{wasserstein_analysis.pgf}
        \end{minipage}
    }%
    {Performance under distribution drift.\label{fig:distribution_shift}}
    {Weekly profit improvement of the \textit{Classification + Policies} model for the HPP (light blue) and wind-only (dark blue) portfolios over the arbitrage-free \textit{Bid Forecast} benchmark across the rolling test windows. At the horizontal dashed line at 0, both models make the same profit. The shaded background bands encode the distribution drift of each test window, measured as the sliced Wasserstein distance between the joint feature-target distribution of the training set and that of the test set in a given window. (a) shows the results for DK1; (b) shows the results for DE/LU.}
\end{figure}
Figure~\ref{fig:distribution_shift} shows the out-of-sample profit improvement for the HPP and wind-only portfolios and both markets, DK1 and DE/LU, over all testing windows given a high weight on the mean profit objective term in (\ref{eq:Optimization_a}), $\beta=0.9$. A positive value reflects the profit that the arbitrage learning adds on top of simply bidding the wind forecast. In addition, the profit improvement is shown in relation to the level of distribution drift between the training and testing set for each window that is visualized as a shaded background band for each window and measured as the sliced Wasserstein distance between the joint feature-target distributions \citep{Bonneel2015,Peyre2019}.

Across all windows, the HPP (light blue) achieves more arbitrage profit than the standalone wind farm (dark blue) in both markets. The reason for this is the additional internal flexibility provided by the electrolyzer. The electrolyzer can absorb part of the mismatch between the day-ahead quantity bid and the realized wind power production internally, which lets the policy place more opportunistic bids with larger magnitude and capture a larger share of the price spread without exposing the portfolio to the full imbalance cost. The HPP improvements therefore peak well above the benchmark (around $+75\%$ in DK1 and above $+100\%$ in DE/LU), whereas the wind-only portfolio stays much closer to the benchmark model and seldom shows a profit improvement of more than about $30\%$.

The distribution drift bands explain much of the window-to-window variation in testing. In windows with small drift, where the test distribution still resembles the training data, the classifier and policies generalize better and the arbitrage learning delivers its largest profit improvement. Therefore, most of the pronounced positive spikes coincide with the lighter bands. As the distribution drift grows, the profit improvement tends to shrink, and under the most severe drifts (darker red bands), it collapses toward zero or even turns negative, meaning the learned policy can do worse than simply bidding the forecasted wind power production. This is the expected failure when the environment is non-stationary, i.e., the model was trained on a different environment of features and target and the confident opportunistic out-of-sample bids are increasingly placed on the wrong side of the spread. This is most evident for the wind-only case in DK1, where the third window with much larger distribution drift than the surrounding testing windows  causes the profit improvement to drop to around $-100\%$.

\begin{figure}
    \FIGURE
    {%
        \begin{minipage}[t]{\textwidth}
            \centering
            \input{profit_distributions.pgf}
        \end{minipage}
    }%
    {Profit distributions across all test windows.\label{fig:profit_distributions}}
        {Distribution of realized profits \textit{per hour} over all 22 rolling test windows for the HPP portfolio in DK1. The models compared are defined in Table~\ref{tab:models}. Each panel marks the mean profit (black, dashed) and the CVaR\textsubscript{5\%} of the profit (red, solid). The horizontal axis is clipped for better visibility.}
\end{figure}

In the following, we limit the analysis to the HPP case and the DK1 market to avoid redundancies. Figure~\ref{fig:profit_distributions} reports the realized per-hour profit distributions over all 22 rolling test windows for the HPP portfolio comparing all models that are defined in Table~\ref{tab:models}. For the \textit{Classification} models (b, d, f), the confidence thresholds are tuned as described in Section~\ref{sec:classif} and $\beta$ is now set to $\beta=0.7$ to balance more between both objective terms in~(\ref{eq:Optimization_a}). The two hindsight-based models (a,b) limit the achievable performance. Our proposed model \textit{Classification + Policies} (f) outperforms the \textit{Bid Forecast} (c) and \textit{Single Policy} (e) benchmark models in terms of mean profit, but at a lower CVaR\textsubscript{5\%} value. The \textit{Classification + Policies} shows a mean profit that is 7\% higher than the \textit{Bid Forecast} and 4\% higher than the \textit{Single Policy} with a lower CVaR\textsubscript{5\%} of 19\% and 51\% respectively. Compared to the \textit{Classification + All-or-Nothing}, it shows a slightly lower mean profit of 3\%, while the CVaR\textsubscript{5\%} is 86\% better. This shows that \textit{All-or-Nothing} is an extreme case of our proposed \textit{Policies} model in Step 2, and therefore, has the heaviest tail of all models but also the highest mean profit. The \textit{Classification + Policies} model, by contrast, retains a comparable mean profit while significantly compressing this tail as its learned policies per class scale the opportunistic bid magnitude with the contextual features instead of committing to the all-or-nothing extremes. Hence, it is capable of finding a balance between a higher mean profit and a higher risk (CVaR\textsubscript{5\%}), which is important especially in non-stationary environments.

\begin{table}[t]
\TABLE
{Error propagation from Step 1 to Step 2 in the testing set due to incorrect classification.\label{tab:error_propagation}}
{\ifdefined\modelcolwidth\else\newlength{\modelcolwidth}\fi
\settowidth{\modelcolwidth}{\begin{tabular}[b]{@{}c@{}}\textit{Bid}\\\textit{Forecast} (1/h)\\(Reference)\\\textit{Classification}\\\textit{+ Policies}\\(Our Model)\\\textit{Hindsight}\\\textit{+ Policies}\\(Oracle\\Classification)\end{tabular}}
\begin{tabular*}{\textwidth}{@{\extracolsep{\fill}}l *{3}{w{c}{\modelcolwidth}}@{}}
\hline\up
 & \begin{tabular}[b]{@{}c@{}}\textit{Bid} \textit{Forecast}\\(Reference)\\(€/h)\end{tabular} & \begin{tabular}[b]{@{}c@{}}\textit{Classification}\\\textit{+ Policies}\\(Our Model)\end{tabular} & \begin{tabular}[b]{@{}c@{}}\textit{Hindsight}\\\textit{+ Policies}\\(Oracle\\Classification)\end{tabular} \\ \hline\up
Hours with correct classification (19.5\%) & $\phantom{}182.9$ & $\phantom{}{+}116.7$\% & $\phantom{}{+}144.0$\% \\
Hours with incorrect classification (19.6\%) & $\phantom{}231.7$ & $\phantom{0}{-}57.3$\% & $\phantom{0}{+}58.0$\%\down \\ \hline\up
All hours & $\phantom{}223.0$ & $\phantom{00}{+}6.9$\% & $\phantom{0}{+}87.6$\%\down \\ \hline
\end{tabular*}}
{\NOTE{Note.}Mean profit per hour over all test windows for the HPP in DK1. The test hours are split by the Step~1 decision of the \textit{Classification + Policies} model, i.e., an opportunistic bid in the correct or in the incorrect direction of the price spread with the share of all test hours in parentheses. All hours also include the remaining testing hours with arbitrage-free bids. All models are evaluated on the same hours. \textit{Bid Forecast} shows the mean profit, while \textit{Classification + Policies} and \textit{Hindsight + Policies} show the change in mean profit relative to \textit{Bid Forecast}. A positive sign indicates a profit improvement, while a negative sign indicates a loss.}
\end{table}

Since Step~2 only chooses the extent of an opportunistic bid in the direction set by Step~1, it cannot reverse a classification error, but only limit its economic cost. To quantify the economic cost that can be attributed to Step~1, we introduce the benchmark \textit{Hindsight + Policies}, which replaces the classifier with the realized sign of the price spread in every hour (oracle classification) and learns the Step~2 policies on these true classes with the same weight $\beta=0.7$. In Table~\ref{tab:error_propagation}, this benchmark is compared with the \textit{Classification + Policies} and \textit{Bid Forecast} models for the HPP in DK1, split by whether the Step~1 classification of the \textit{Classification + Policies} model is correct or incorrect in each test hour. In correctly classified hours, \textit{Classification + Policies} gains 117\% in profit compared to the arbitrage-free \textit{Bid Forecast} benchmark. There remains a gap to the oracle classification model, as the policies in Step 2 have been trained on different training sets. In incorrectly classified hours, the opportunistic bid is placed on the wrong side of the spread. There, \textit{Classification + Policies} loses 57\% in profit relative to the arbitrage-free bid, whereas \textit{Hindsight + Policies} still gains 58\%. Hence, the direction error in Step~1 propagates through the model to the decision. However, the profit improvement of correctly classified hours outweighs the loss of incorrectly classified testing hours, yielding a net improvement in profit of 7\% over all hours for our model. Note that all hours include the arbitrage-free testing samples, which are identical for the \textit{Bid Forecast} and \textit{Classification + Policies} models, in addition to the correctly and incorrectly classified testing sets. The resulting gap between \textit{Classification + Policies} and \textit{Hindsight + Policies} of around 80\% profit improvement over the \textit{Bid Forecast} can be attributed to Step~1. It accounts for about 65\% of the total gap of about €275/h to the \textit{Hindsight} benchmark in Figure~\ref{fig:profit_distributions}a. The remaining 35\% can be attributed to Step~2. This aligns with the alternative decomposition via \textit{Classification + Hindsight} in Figure~\ref{fig:profit_distributions}b, which attributes €179/h (65\%) to the classification in Step~1 and €96/h (35\%) to Step~2. Hence, both oracle decompositions attribute the same shares of the total gap to the two steps.

\begin{table}[t]
\TABLE
{Impact of the residual load forecast error used as a feature on resulting profit in the testing set.\label{tab:forecast_error}}
{\ifdefined\modelcolwidth\else\newlength{\modelcolwidth}\fi
\settowidth{\modelcolwidth}{\begin{tabular}[b]{@{}c@{}}\textit{Bid}\\\textit{Forecast}\\(Reference)\\\textit{Classification}\\\textit{+ Policies}\\(Our Model)\\\textit{Hindsight}\\\textit{+ Policies}\\(Oracle\\Classification)\end{tabular}}
\begin{tabular*}{\textwidth}{@{\extracolsep{\fill}}l *{3}{w{c}{\modelcolwidth}}@{}}
\hline\up
 & \begin{tabular}[b]{@{}c@{}}\textit{Bid} \textit{Forecast}\\(Reference)\\(€/h)\end{tabular} & \begin{tabular}[b]{@{}c@{}}\textit{Classification}\\\textit{+ Policies}\\(Our Model)\end{tabular} & \begin{tabular}[b]{@{}c@{}}\textit{Hindsight}\\\textit{+ Policies}\\(Oracle\\Classification)\end{tabular} \\ \hline\up
Hours with a comparatively lower error & $\phantom{}219.5$& $\phantom{}{+}9.2$\% & $\phantom{}{+}96.1$\% \\
Hours with a comparatively higher error & $\phantom{}226.5$ & $\phantom{}{+}4.7$\% & $\phantom{}{+}79.3$\%\down \\ \hline\up
All hours & $\phantom{}223.0$ & $\phantom{}{+}6.9$\% & $\phantom{}{+}87.6$\%\down \\ \hline
\end{tabular*}}
{\NOTE{Note.}Mean profit per hour over all test windows for the HPP in DK1. The test hours are split at the median absolute forecast error $|\varepsilon_\tau|=323~\mathrm{MW}$ of the residual load into two sets of hours with a lower and a higher residual load forecast error. The union of these two sets is the set of all test hours. All models are evaluated on the same hours. \textit{Bid Forecast} shows the mean profit, while \textit{Classification + Policies} and \textit{Hindsight + Policies} show the change in mean profit relative to \textit{Bid Forecast}. A positive sign indicates a profit improvement, while a negative sign indicates a loss.}
\end{table}

In addition, the classifier in Step~1 relies on forecasted features, e.g., day-ahead forecasts of load, wind and solar generation, which can be inaccurate themselves. Hence, we analyze how the errors of the forecasted features impact our  model testing performance by splitting the test hours at the median of the absolute residual load forecast error $\varepsilon_\tau$ (323~MW), i.e., the realized minus the forecasted residual load:
\begin{equation}
    \varepsilon_\tau = R_\tau - \hat{R}_\tau,
    \quad \text{with} \quad
    R_\tau = \mathrm{Load}_\tau - \mathrm{Wind}_\tau - \mathrm{Solar}_\tau,
\end{equation}
where $\mathrm{Load}_\tau$, $\mathrm{Wind}_\tau$ (onshore and offshore) and $\mathrm{Solar}_\tau$ are the realized load and generation in DK1, and $\hat{R}_\tau$ is the forecasted residual load. The residual load is a key driver of the system imbalance and thus of our classification target. Note that the day-ahead wind forecasts are on a smaller scale than the realized generation, so each forecast (load, onshore wind, offshore wind and solar) is first recalibrated linearly to its realization by least squares on the training data of the respective rolling window. Hence, $\varepsilon_\tau$ measures the deviation from the relation between forecast and realization that the classifier can learn from its training data. In Table~\ref{tab:forecast_error}, we compare the testing hours with a comparatively higher residual load forecast error, i.e., $|\varepsilon_\tau|\geq323~\rm{MW}$, with the hours with comparatively lower error, i.e.,  $|\varepsilon_\tau|<323~\rm{MW}$. It shows that the profit improvement of our model \textit{Classification + Policies} shrinks from 9\% in hours with lower error to 5\% in hours with higher error. This is a substantially larger relative decrease compared to the oracle classification model \textit{Hindsight + Policies} that shrinks profit gains from 96\% to 79\%. The additional decrease of \textit{Classification + Policies} can therefore be attributed to increasing prediction difficulty in Step~1 with less accurate forecasts as features, i.e., inaccurate forecasted features decrease classification performance and therefore impact the bidding decisions in Step~2. Overall, Step~1 is a major source of economic loss. Improving the classifier, e.g., by a decision-focused loss or online updating under distribution drift, is therefore a promising way to close the gap to the \textit{Hindsight} benchmark.


\subsection{Illustration of Opportunistic Bids}\label{sec:results_bids}
\begin{figure}
    \FIGURE
    {%
        \begin{minipage}[t]{\textwidth}
            \centering
            \input{bids.pgf}
        \end{minipage}
    }%
    {Bidding behavior.\label{fig:bids}}
    {Typical bidding behavior of the \textit{Classification + Policies} model in the testing phase for the HPP. The top three panels show the consecutive day-ahead bids $p^{\rm DA}$ (blue) and the forecasted wind power production $\hat{P}^{\rm W}$ (black) for three weights $\beta$ in \eqref{eq:Optimization_a} over one testing week (October 13-19,~2025). The shaded area marks opportunistic long bids (green) and opportunistic short bids (red). For arbitrage-free bids, the day-ahead bid is identical to the forecasted wind power production. The bottom panel shows the electrolyzer power consumption $p^{\rm H}$ of the HPP for the same period.}
\end{figure}
Figure~\ref{fig:bids} illustrates in the upper three panels the typical bidding behavior of our proposed \textit{Classification + Policies} model over a representative test period for the HPP (blue line) in DK1. Whenever the classifier is not confident enough in the sign of the price spread, the policy falls back to the arbitrage-free bid and bids at a quantity identical to the wind power forecast $\hat{P}^{\rm W}_\tau$ (black line). When the classifier is sufficiently confident, the policy instead places an opportunistic bid that differs from the forecast. An opportunistic long bid offers a larger power quantity to the day-ahead market than is expected to be produced (green shaded area), anticipating that the balancing price will settle below the day-ahead price, and saturates at the maximum day-ahead capacity $\overline{P}^{\rm W}=7.2$~MW. An opportunistic short bid places a quantity smaller than the forecast (red shaded area) to make arbitrage profit on the balancing market. It saturates at the maximum capacity that can be bought in the day-ahead market, which is $-\overline{P}^{\rm H}=-3.6$~MW for the HPP, providing additional flexibility compared to a standalone wind farm that can only offer non-negative bids. Hence, the HPP can make larger short opportunistic bidding decisions to increase the profits.

The bids are shown for three weights $\beta$ in~(\ref{eq:Optimization_a}), which illustrate the effect of changing the weighting between mean profit and CVaR of the profit in the objective function. For $\beta=1$, the policies maximize the mean profit only, which is linear in $p_\tau^{\rm DA}$. The hindsight optimum therefore still lies at a bound, and the opportunistic bids frequently saturate at $\overline{P}^{\rm W}$ or $-\overline{P}^{\rm H}$. For $\beta<1$, the CVaR term in~(\ref{eq:Optimization_a}) can lead to reduced deviations $z_\tau$ in~(\ref{eq:policy}) that improve the tail of the profit distribution of the training set. Hence, the bids for $\beta=0.5$ and $\beta=0$ increasingly remain between the forecast and the bounds, and for $\beta=0$ they rarely reach a bound. A smaller $\beta$, therefore, shrinks the magnitude of the opportunistic deviations $z_\tau$. The electrolyzer dispatch $p_\tau^{\rm H}$ in the bottom panel is identical in all three cases, as the balancing position absorbs any residual, so the dispatch~(\ref{eq:hpp_balancing}) depends only on the balancing and hydrogen prices, not on $\beta$.

\subsection{Sensitivity Analysis on Balancing Price Forecast Error}\label{sec:results_balancing_sensitivity}
\begin{figure}
    \FIGURE
    {%
        \begin{minipage}[t]{\textwidth}
            \centering
            \input{balancing_forecast_sensitivity.pgf}
        \end{minipage}
    }%
    {Sensitivity to the balancing price forecast error.\label{fig:balancing_sensitivity}}
    {Mean profit per hour of the \textit{Classification + Policies} model in DK1 over all test windows as a function of the quality of the balancing price forecast $\hat{\lambda}^{\rm B}_\tau$ that enters the electrolyzer dispatch~\eqref{eq:hpp_balancing}, measured as the mean absolute error (MAE) with respect to the realized balancing price. Shown are two different methods used to generate the forecast errors: the blue line with 90\% confidence interval adds i.i.d.\ Student-$t$ noise with increasing scale to the realized balancing price, while the red line scales the error of a naive lag-1h forecast by a factor $w > 0$, where $w=1$ is the naive forecast itself. The wind-only portfolio is given by the horizontal dashed gray line.}
\end{figure}
The electrolyzer dispatch~(\ref{eq:hpp_balancing}) is optimized against a forecast $\hat{\lambda}^{\rm B}_\tau$ of the balancing price, while the balancing volume is settled ex-post at the realized price $\lambda^{\rm B}_\tau$. The previously reported results assume perfect foresight of the balancing price in real time, i.e., $\hat{\lambda}^{\rm B}_\tau = \lambda^{\rm B}_\tau$. However, this is an optimistic assumption given the volatility of balancing prices documented in Section~\ref{sec:preliminaries}. To quantify how much of the flexibility value the HPP still yields under imperfect forecasts compared to the wind-only portfolio, we degrade the balancing price forecast while keeping the day-ahead bids $p^{\rm DA}_\tau$, the realized wind power production, and the realized day-ahead and balancing prices fixed. Hence, the forecast error propagates to the profit only through the electrolyzer dispatch and the resulting balancing volume. We consider two methods to generate the forecast errors: (i) adding i.i.d.\ Student-$t$ noise with $\nu=3$ degrees of freedom and increasing scale to the realized price, representing a conservative case, and (ii) scaling the error of a naive lag-1h forecast, $\lambda^{\rm B}_{\tau-1} - \lambda^{\rm B}_{\tau}$, by a factor $w > 0$, which inherits the temporal structure of the prices, i.e., the error is large exactly in the hours where the price moves sharply and small in stable hours. In case of $w = 1$, the forecast is exactly the lag-1h price realization. Both forecast errors are indexed by the resulting mean absolute error (MAE), so that the two curves can be compared at equal forecast accuracy.

Figure~\ref{fig:balancing_sensitivity} shows the resulting mean profit per hour over the whole testing period for the HPP in DK1. Under perfect foresight, the HPP earns €238.3/h, which is 8.4\% more than the wind-only portfolio at €219.9/h (gray dashed line). This profit surplus decreases as the forecast quality degrades and vanishes at an MAE of €45.9/MWh for the i.i.d.\ Student-$t$ noise (blue line) and €60.2/MWh for the scaled lag-1h error (red line). The naive lag-1h forecast itself has an MAE of €61.7/MWh and yields €215.8/h, i.e., 1.9\% below the wind-only portfolio. Up to an MAE of around €55/MWh, the scaled lag-1h error decreases the profit much more slowly than the i.i.d. noise, after which it drops sharply. The reason is that its errors concentrate in a few price spikes while the median error stays small, whereas the i.i.d.\ noise affects each hour randomly and can therefore flip the dispatch decision in hours in which the balancing price is close to the hydrogen price (€2/kg). Hence, the HPP outperforms the wind-only portfolio up to a balancing price forecast error of around €46/MWh to €60/MWh, depending on the error model.

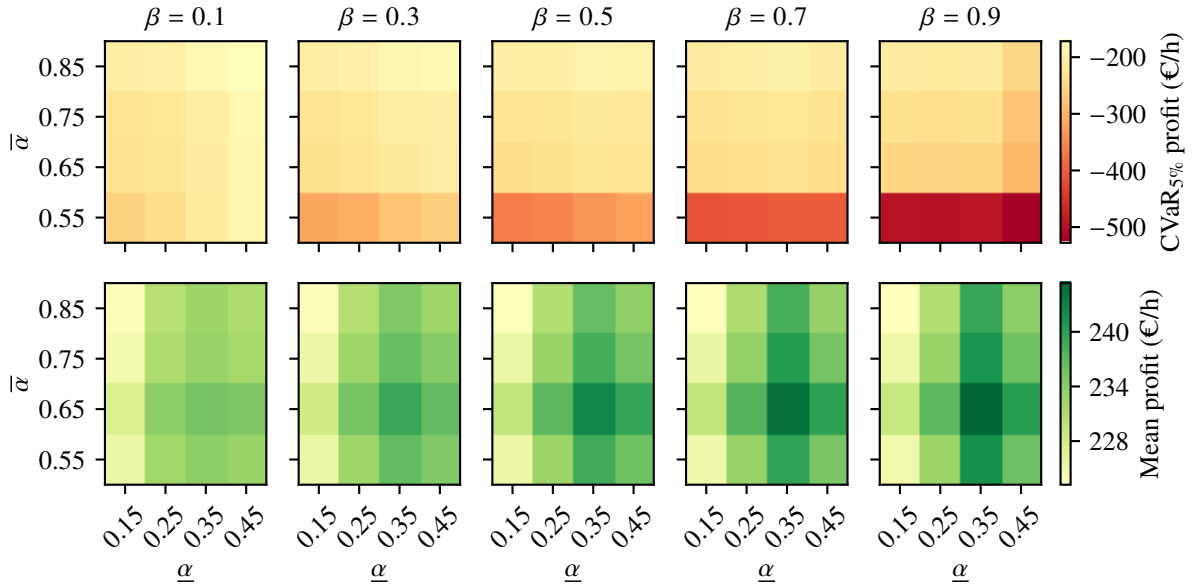
\subsection{Risk Sensitivity Analysis}\label{sec:risk_sensitivity}
\begin{figure}
    \FIGURE
    {%
        \begin{minipage}[t]{\textwidth}
            \centering
            \input{risk_sensitivity_3d.pgf}
        \end{minipage}
    }%
    {Risk sensitivity analysis.\label{fig:risk_sensitivity}}
    {Shown are (top row) the CVaR\textsubscript{5\%} of the profit and (bottom row) the mean profit of the testing phase for the \textit{Classification + Policies} model, as a function of the two confidence thresholds $\underline{\alpha}$ and $\overline{\alpha}$, for five values of $\beta$ in the optimization objective function (\ref{eq:Optimization_a}). The weight $\beta$ provides a trade-off between mean profit and CVaR\textsubscript{5\%} in (\ref{eq:Optimization_a})  ($\beta=1$ maximizes only mean profit, $\beta=0$ only CVaR\textsubscript{5\%}). The confidence thresholds are given as exogenous parameters and are not tuned here.}
\end{figure}

In our framework, two groups of parameters tune the risk exposure of the day-ahead bidding decision and they act on different stages of the pipeline. The first group is the two confidence thresholds $\underline{\alpha} \in [0,0.5]$ and $\overline{\alpha} \in [0.5,1]$ of the classification step, which set the classifier confidence required before an opportunistic bid is placed through the threshold rule~\eqref{eq:threshold_rule}. They set the risk of making an opportunistic bid, i.e., when and in what direction the trader deviates from the arbitrage-free bid. Lowering $\underline{\alpha}$ toward $0$ and raising $\overline{\alpha}$ toward $1$ widen the arbitrage-free band and yield fewer, more selective and confident opportunistic bids, whereas pushing both thresholds toward $0.5$ admits more opportunistic bidding with less confidence. The second group is the single weight $\beta \in [0,1]$ in the policy objective function~\eqref{eq:Optimization_a}, which sets the risk of the magnitude of an opportunistic bid. When $\beta = 1$, the objective is to maximize mean profit solely, while decreasing $\beta$ shifts weight onto CVaR\textsubscript{5\%} of the profit and shrinks the tail.

Figure~\ref{fig:risk_sensitivity} reports the joint out-of-sample effect of both groups for the \textit{Classification + Policies} model on the HPP portfolio in DK1, showing the CVaR\textsubscript{5\%} of the profit over all testing windows (top row) and the mean profit (bottom row) over the $\underline{\alpha} \times \overline{\alpha}$ grid for different values of $\beta$. The figure shows that both the mean profit and CVaR\textsubscript{5\%} experience an asymmetric effect of the confidence thresholds, which is, however, larger for the CVaR\textsubscript{5\%} value. In particular, the upper threshold $\overline{\alpha}$ appears to be a dominant factor improving CVaR\textsubscript{5\%} consistently across all values of $\beta$. The reason for this is that opportunistic long bids, which offer a larger quantity than is forecasted to be produced, are the main source of large imbalance costs when the spread realizes in the opposite direction, and a higher $\overline{\alpha}$ filters out the least confident of these opportunistic long bids. The lower threshold $\underline{\alpha}$ behaves in the opposite way, i.e., allowing more opportunistic short bids with lower confidence (larger $\underline{\alpha}$) tends to improve rather than worsen the tail. The mean profit, by contrast, is maximized at a moderate threshold pair (around $\underline{\alpha}=0.35$ and $\overline{\alpha}=0.65$) rather than at either corner. Therefore, the most arbitrage-free configuration misses profitable arbitrage and lowers the mean profit, while the most opportunistic bidding with least confidence configuration adds little mean profit at a significantly worse tail.

The weight $\beta$ in the optimization objective function~(\ref{eq:Optimization_a}) shows the mean profit--risk trade-off along the columns of Figure~\ref{fig:risk_sensitivity}. Increasing $\beta$ raises the mean profit but deepens the tail, i.e., the CVaR\textsubscript{5\%} becomes more negative. However, this trade-off is strongly asymmetric over the threshold grid. It is mild in the conservative region with few opportunistic bids (top left corner) but severe when many opportunistic long bids with low confidence are admitted (low $\overline{\alpha}$), where the worst-case loss roughly doubles between the most risk-averse ($\beta=0.1$) and the most profit-seeking ($\beta=0.9$) policy. Hence, the weight $\beta$ matters most precisely when the thresholds allow many opportunistic long bids with low confidence, so the two parameter groups interact rather than act independently.

\section{Conclusion}\label{sec:Conclusion}
In this paper, we studied how a price-taking stochastic energy generator should engage in opportunistic arbitrage between the day-ahead and balancing markets under a single-price imbalance settlement. We proposed a \textit{predict-then-contextual-optimize} framework that decomposes the day-ahead bidding decision into three explainable stages to decide when to engage in arbitrage, in what direction, and to what extent. The first two stages are answered in a first step by a probabilistic classification with two confidence thresholds. A linear decision policy per class is learned in the second step through contextual optimization to determine the magnitude of the deviation from the forecast answering the third stage. The proposed model was tested in two European bidding zones (DK1 and DE/LU) for a standalone wind farm and a co-located wind farm and electrolyzer. It showed an increased mean profit compared to the benchmark models. In particular, the co-located electrolyzer substantially raised arbitrage profits by providing additional bidding flexibility. The risk sensitivity analysis showed that the two parameter groups, the two confidence thresholds and the single CVaR weight, allow explainable tuning of the profit--risk trade-off, while the distribution drift analysis confirmed that the arbitrage profit improvements concentrate in windows with small drift and decrease under strong distribution drift.

Our study has several limitations that are left for future research. First, the distribution-drift analysis shows that non-stationarity can limit the arbitrage profits achieved by the proposed framework. 
In the terminology of selective classification, the confidence thresholds implement ambiguity rejection, i.e., abstention close to the decision boundary. The losses observed ex-post under strong distribution drift suggest complementing them with novelty rejection, i.e., abstention when the features lie far from the training distribution~\citep{Hendrickx2024}, based on an out-of-distribution score computed from the features alone.
In addition, an online prediction and policy updating scheme could allow the framework to adapt continuously to changing market conditions.
Second, the proposed framework of two sequential steps forgoes the value of a full co-optimization. Future research should explore how to embed the classification step into the contextual optimization problem to jointly optimize both steps and potentially improve the overall performance.
Third, this work focuses on the perspective of a single price-taking trader. Studying the market-level implications of opportunistic arbitrage bidding, including its effects on liquidity, price convergence, and social welfare when such strategies are adopted at scale, remains an important direction for future research.


\clearpage 

%
\bibliography{science_template} 
\bibliographystyle{plainnat} 

%
%
%
%
%
%


\section*{Acknowledgments}
We gratefully acknowledge the Danish Energy Technology Development and Demonstration Programme (EUDP) for supporting this research through the ViPES2X project (Grant number: 640222-496237), and the Innovation Fund Denmark for supporting our work through the PtX Markets project (Grant number: 150-00001B). We are also grateful to Enfor for providing data and Jan Leisbrock and David Miles-Skov for valuable discussions throughout the course of this work.
\appendix
\numberwithin{table}{section}
\numberwithin{figure}{section}

\section{Feature Overview}\label{app:features}

Table~\ref{tab:app_features} summarizes the feature categories considered in the case study (Section~\ref{sec:case_studies}). The full dataset with all features can be found in~\citet{Heiser2026Code}. All features are z-score normalized, and feature selection is performed on each rolling window using SHAP importance values.

\begin{table}[h!]
\TABLE
{Feature categories considered in the case study.\label{tab:app_features}}
{\begin{tabular}{@{}p{0.24\textwidth} p{0.68\textwidth}@{}}
\hline\up
Category & Features \\ \hline\up
Temporal & Cyclic sin/cos encodings of hour-of-day and day-of-week. \\
Market prices and imbalance volumes & Lagged day-ahead prices, balancing prices, and balancing demand for DK1, DE/LU, and their neighboring zones as well as the lagged continuous price spread. \\
Reserve market & Demand, procurement volumes, and capacity prices for both automatic (aFRR) and manual frequency restoration reserve (mFRR) across all zones. \\
Generation and load forecasts & Renewable generation forecasts (wind offshore/onshore, solar) and total generation and load forecasts for all zones. \\
Cross-border exchange & Lagged physical cross-border flows, scheduled exchanges, offered capacities, net transfer forecasts, and net positions between DK1, DE/LU and all neighboring zones. \\
Meteorological & Temperature, cloud cover, wind speed, and wind direction at multiple grid points over DK1 and DE/LU.\\
Feature engineering & Features engineered out of other features, e.g., forecasted residual load and energy balance, mFRR and aFRR volume and price spreads. \down \\ \hline
\end{tabular}}{}
\end{table}

\section{Classification Model and Hyperparameter Grids}\label{app:models}

This appendix lists the LGBM classification model evaluated in Section~\ref{sec:case_studies}, including its fixed parameters and the hyperparameter grids searched on the validation set. The model uses balanced class weights to counteract class imbalance.


\begin{table}[h!]
\TABLE
{LightGBM (LGBM) hyperparameter grid.\label{tab:app_lgbm}}
{\begin{tabular}{@{}l@{\quad}l@{\quad}l@{}}
\hline\up
Parameter & Fixed value & Tuned grid \\ \hline\up
No.\ estimators & 500 & --- \\
Row subsampling & 0.7 & --- \\
Feature fraction & 0.7 & --- \\
$\ell_1$ regularization ($\alpha$) & 0.1 & --- \\
$\ell_2$ regularization ($\lambda$) & 10 & --- \\
Min.\ split gain & 0.5 & --- \\
Learning rate & --- & $\{0.01,\;0.03\}$ \\
No.\ leaves & --- & $\{7,\;15\}$ \\
Min.\ child samples & --- & $\{20,\;50\}$ \\
Max.\ depth & --- & $\{2,\;3\}$\down \\ \hline
\end{tabular}}{}
\end{table}


\end{document}